\pdfoutput=1
\documentclass[acmsmall,screen,nonacm]{acmart}
\usepackage{mathtools}
\usepackage{amsmath}
\usepackage{amsthm}
\usepackage{mathpartir}
\usepackage{listings}
\usepackage{booktabs}
\usepackage{makecell}
\usepackage{xspace}
\usepackage{stmaryrd}
\usepackage{pifont}
\usepackage{multirow}
\usepackage{wrapfig}
\usepackage{bussproofs}
\usepackage[inline]{enumitem}
\usepackage{enumitem}
\usepackage{colortbl}
\usepackage[utf8]{inputenc}
\usepackage[T1]{fontenc}
\usepackage{microtype}
\usepackage{dirtree}

\usepackage{aliascnt}
\newtheoremstyle{thmbasic}
  {2pt}{2pt}               
  {\itshape}               
  {}                       
  {\bfseries\itshape}      
  {.}                      
  { }                      
  {}                       
\newtheoremstyle{defbasic}
  {2pt}{2pt}               
  {\normalfont}            
  {}                       
  {\bfseries\itshape}      
  {.}                      
  { }                      
  {}                       
\theoremstyle{thmbasic}
\newtheorem{theorem}{Theorem}[section]
\newaliascnt{lemma}{theorem}
\newtheorem{lemma}[lemma]{Lemma}
\aliascntresetthe{lemma}

\theoremstyle{defbasic}
\newaliascnt{definition}{theorem}
\newtheorem{definition}[definition]{Definition}
\aliascntresetthe{definition}

\newtheoremstyle{examplebasic}
  {3pt}{3pt}
  {\normalfont}
  {}
  {\itshape}
  {:}
  { }
  {}
\theoremstyle{examplebasic}

\usepackage{xr-hyper}
\usepackage{hyperref}
\usepackage{cleveref}
\crefname{section}{\S}{\S}
\crefname{Section}{\S}{\S}
\crefname{theorem}{Theorem}{Theorems}
\Crefname{theorem}{Theorem}{Theorems}
\crefname{lemma}{Lemma}{Lemmas}
\Crefname{lemma}{Lemma}{Lemmas}
\crefname{definition}{Definition}{Definitions}
\Crefname{definition}{Definition}{Definitions}
\definecolor{lightTeal}{RGB}{224, 247, 250}
\definecolor{lightCream}{rgb}{1.0, 0.98, 0.94}
\definecolor{mediumTeal}{RGB}{200, 235, 238}  
\definecolor{mediumCream}{RGB}{255, 248, 220} 
\definecolor{softBlue}{RGB}{218, 232, 252}    
\definecolor{paleYellow}{RGB}{255, 255, 224}  
\definecolor{paleGreen}{RGB}{230, 255, 230}   
\definecolor{deepCream}{RGB}{255, 240, 200}
\definecolor{warmCream}{RGB}{255, 235, 180}
\definecolor{goldCream}{RGB}{255, 230, 170}
\newcommand{\nat}{\mathsf{N}}

\newcommand{\mypara}[1]{\smallskip\noindent\textbf{{#1}.}}

\newcommand{\tool}{\textsc{SymFC}\xspace}

\newcommand{\ife}[3]{\textcolor{purple}{if}~#1~\textcolor{purple}{then}~#2~\textcolor{purple}{else}~#3}

\newcommand{\mono}[1]{\ensuremath{\mathtt{#1}}}
\newcommand{\m}[1]{\mono{#1}}

\newcommand{\vect}[1]{\overrightarrow{\mono{#1\;}}}

\newcommand{\symrule}[1]{\colorbox{symrulebg}{$#1$}}

\newcommand{\axiomc}[1]{\AxiomC{\mono{#1}}}
\newcommand{\unaryc}[1]{\UnaryInfC{\mono{#1}}}
\newcommand{\binaryc}[1]{\BinaryInfC{\mono{#1}}}

\newcommand{\rightlabel}[1]{\RightLabel{\mono{#1}}}

\definecolor{symrulebg}{gray}{0.92}
\newenvironment{symprooftree}%
  {\begin{center}\proofSkipAmount \leavevmode}%
  {\colorbox{symrulebg}{\DisplayProof}\proofSkipAmount \end{center}}

\newcommand{\app}[2]{#1 \ #2}
\newcommand{\abs}[2]{\lambda #1 . \ #2}
\newcommand{\quant}[2]{\forall #1 . \ #2}
\newcommand{\cast}[2]{#1 \triangleright #2}
\newcommand{\ite}[3]{\textcolor{purple}{if} \; #1 \; \textcolor{purple}{then} \; #2 \; \textcolor{purple}{else} \; #3}
\newcommand{\case}[2]{\textcolor{purple}{case} \ #1 \ \textcolor{purple}{of} \ #2}

\newcommand{\alt}[3]{\app{#1}{#2} \to #3}
\newcommand{\alts}[1]{\vect{#1}}

\newcommand{\bigstep}[4]{#1; #2 \vdash #3 \Downarrow #4}
\newcommand{\bigstepb}[5]{#1; #2 \vdash #4 \Downarrow^{#3} #5}
\newcommand{\constep}[3]{#1 \vdash #2 \Downarrow #3}
\newcommand{\instance}[3]{#1 \sqsupseteq_{#2} #3}

\newcommand{\foldb}[6]{#1 ; #2 ; #5 \vdash #4 \Downarrow^{#3}_{case} #6}
\newcommand{\applyb}[5]{#1 \vdash \app{#3}{#4} \Downarrow^{#2}_{apply} #5}

\newcommand{\forceb}[4]{#1 \vdash #3 \Downarrow^{#2}_{thunk} #4}

\newcommand{\reads}[2]{\llbracket #1 \rrbracket\if\relax\detokenize{#2}\relax\else_{#2}\fi}

\newcommand{\thunk}[2]{(#1, #2)}
\newcommand{\budget}{\bot_{k}}

\newcommand{\unreachable}{\emptyset}
\newcommand{\ub}{\mathord{?}}
\newcommand{\subst}{\mathrm{\Gamma}}
\newcommand{\pathc}{\mathrm{\Phi}}
\newcommand{\prim}{\otimes}
\newcommand{\coInst}{\gamma}
\newcommand{\tyInst}{\tau}

\newcommand{\rcoercion}{\sim_R}
\newcommand{\ncoercion}{\sim_N}

\newcommand{\embed}{\mathrm{\Psi}}

\newcommand{\ie}{\emph{i.e.},\xspace}
\newcommand{\eg}{\emph{e.g.},\xspace}

\def\emphbf#1{\textbf{\emph{#1}}} \def\spmid{\ \mid \ }
\def\bnf{\vcentcolon\vcentcolon=}

\newcommand{\sys}{\textsc{SymFC}\xspace}
\newcommand{\calc}{\textsc{SymCore}\xspace}
\newcommand{\core}{\textsc{ConCore}\xspace}
\newcommand{\clash}{C$\lambda$aSH\xspace}

\newcommand{\FC}{System FC\xspace}

\newcommand{\appCref}[1]{%
  \begingroup
    \Crefname{section}{Appendix}{Appendices}%
    \Cref{#1}%
  \endgroup
}
\newcommand{\appcref}[1]{%
  \begingroup
    \crefname{section}{appendix}{appendices}%
    \cref{#1}%
  \endgroup
}

\makeatletter
\newcommand{\dropImportedTocIndents}{%
  \@for\level:=-1,0,1,2,3,4,5\do{%
    \expandafter\global\expandafter\let\csname r@tocindent\level\endcsname\@undefined}}
\makeatother

\newcommand{\cmark}{\ding{51}}
\newcommand{\xmark}{\ding{55}}

\begin{document}

\title{\tool: Configurable Semantics for All Paths Symbolic Execution in Haskell via Embeddings}
\author{Robin Webbers}
\orcid{0009-0000-3493-837X}
\affiliation{%
  \institution{Vrije Universiteit Amsterdam}
  \city{Amsterdam}
  \country{The Netherlands}
}
\email{r.m.a.webbers@vu.nl}

\author{Wind Wong}
\orcid{0009-0005-8306-7968}
\affiliation{%
  \institution{Vrije Universiteit Amsterdam}
  \city{Amsterdam}
  \country{The Netherlands}
}
\email{t.f.w.wong@vu.nl}

\author{Klaus von Gleissenthall}
\orcid{0000-0003-0826-4425}
\affiliation{%
  \institution{Vrije Universiteit Amsterdam}
  \city{Amsterdam}
  \country{The Netherlands}
}
\email{k.freiherrvongleissenthal@vu.nl}


\begin{abstract}
Lightweight methods like QuickCheck have proven to be successful in functional
languages and beyond, as they let users specify properties as Boolean functions
directly in code.
We can replicate some of their benefits for (bounded) proofs via symbolic
execution. If we can execute arbitrary functions symbolically, we can specify
properties just as easily.
Unfortunately, existing symbolic execution tools for Haskell are restricted to
a subset of the language that excludes many real-world applications like IO;
they also do not scale well to larger programs, making them a poor fit
for general purpose lightweight verification.
One reason behind these limitations is that existing techniques only offer fixed
symbolic semantics for SMT base types. This makes it hard to cover more
programming constructs and libraries; it also makes it difficult to fine-tune
symbolic representations for better performance.

We present \sys, a new symbolic execution method for Haskell that offers
configurable semantics via embeddings.
An embedding of a function in \sys is just another Haskell function that can use
arbitrary code, as well as SMT primitives.
Embeddings make it easy to support new programming constructs and fine-tune
symbolic representations for speed.
To integrate embeddings into Haskell's type system, we make use of Haskell's
coercions, which let us transparently convert between a Haskell type and the
type of its embedding.
We realize embeddings in \tool, an all-paths symbolic execution engine for
Haskell, which lets us symbolically represent arbitrary types and values in GHC
core.
We implement \tool as a GHC plugin, and evaluate it by checking properties of
Haskell's container library, and the \texttt{iproute} library.
We build embeddings for \texttt{IORef}, Haskell's
\texttt{Ptr}/\texttt{ForeignPtr} interface, and the hardware library \clash,
and our evaluation reports a 6.25x speedup on G2's own benchmarks.
\end{abstract}

\maketitle
\section{Introduction}
\mypara{Motivation and Problem}
Lightweight formal methods like Property Based Testing (PBT) \`a la
QuickCheck~\cite{quickcheck} is
among the most widely used formal tools in functional languages
and beyond~\cite{10.1145/3597503.3639581}.
What makes PBT so appealing is its ease of use: users specify
the properties they want to test directly as Boolean functions as part of
the code; checks are fast since they only need to run the underlying program.
Symbolic execution~\cite{king} can offer the same interface with stronger
guarantees: given a property written as a Boolean function, it considers all
inputs at once, and either proves the property for all of them, up to a recursion bound,
or returns a counterexample.
Unlike PBT, which simply runs compiled code, this only works if the symbolic
executor can give meaning to the property and every function it depends on.
But symbolically executing real Haskell is difficult. Every GHC Core program
reduces to primitive operations---arithmetic on unboxed types such as
\lstinline{Int64#}, mutable references threaded through the \lstinline{State# RealWorld} 
token, raw memory accessed through \lstinline{Addr#}, and foreign
calls---each in need of a symbolic meaning.
GHC 9.2 alone exports about 1300 such primitives defined over around 130 primitive
types.
Existing tools like G2~\cite{lazycounter} hardwire this meaning into their
engine: each primitive gets (at best) a single fixed model, (mapping, say,
\texttt{Int} to SMT BitVectors), which is then lifted to the rest of the
language.
But this causes two problems.
First, code that reaches a primitive without a model cannot be executed
symbolically until the engine's developers add one.
Second, there is at most one encoding per primitive, but the right encoding may
depend on the property we want to verify.
For example, client code that uses \texttt{text}, will want to model strings as
SMT strings, which is much faster than a bit-level encoding; but to verify the
implementation of \texttt{text} itself, we need its bit-level operations and
pointers.
We therefore need program semantics that are \emph{configurable} by the user, on
a per-property basis.

\mypara{Our Solution}
In this work, we present \tool a new symbolic execution method for Haskell that
offers \emph{configurable} semantics via a new notion of embeddings.
An embedding in \tool replaces a type's representation or a function's
implementation with ordinary Haskell code, which may use arbitrary Haskell as
well as SMT primitives from any theory.
Embeddings allow us to fine-tune symbolic representations according to need on a
per-property level which helps both coverage and scalability.
We integrate embeddings into Haskell through coercions: a type embedding yields
a coercion axiom that can convert between a type and its interpretation, and
GHC's type checker verifies that every term embedding is consistent with the
type embeddings it uses.
Because embeddings are compiled to GHC Core and run by the same evaluator as
the program, there is no separate modelling language, and embeddings can be
executed and tested like any other Haskell code.
Reusing GHC's infrastructure keeps the extension to $\sim${} 500 lines of
Haskell.

To be scalable, \tool performs all-paths symbolic evaluation with
merging~\cite{Rosette,Grisette} directly on GHC Core.
Existing symbolic execution tools for Haskell like G2~\cite{lazycounter} explore
one path at a time: a symbolic value of an algebraic data type is split into one
path per constructor, and only constraints over base types reach the solver.
In \tool, a symbolic value of any type is instead a single merged value: when
a case split or an application meets a symbolic branch, both alternatives are
evaluated under their respective guards and the results are merged back into
one value, so the code that follows runs once rather than once per path.
Unlike Grisette and Rosette, which merge values for code written against a
library or a lifted language subset, \tool does so for unmodified, lazy
Haskell.

We formalise our symbolic execution engine in \calc, a calculus for lazy
all-paths symbolic evaluation of System FC.
Symbolic evaluation in \calc is ordinary evaluation, extended with rules for
symbolic values and merging: our rules represent paths, merged values, failed
assertions, and assumptions uniformely as values inside the calculus, rather
than in terms of a separate store for paths and constraints.
We prove in Rocq that symbolic evaluation is sound and bounded-complete with
respect to concrete evaluation.
Finally, we formalise embeddings as a translation on \core programs that is
based on coercion axioms.

We implement \tool as a GHC plugin and evaluate it on 54 properties, 13 of
them deliberately invalid, over primitive types, Haskell's \texttt{containers}
library and the \texttt{iproute} library, where we find an unchecked
arithmetic overflow on out-of-range mask lengths.
We also give three case studies for embeddings: \textit{IORef}/\textit{IO} to model mutable data,
Haskell's pointer and foreign pointer interface
\textit{Ptr}/\textit{ForeignPtr}, and \clash, a library for writing hardware in
Haskell.
The \clash embedding scales to a realistic design: \tool symbolically executes
AIMCore, an 800-line RISC-V processor, in 11.5 seconds.
Finally, we evaluate \sys against G2 on G2's own benchmarks and report a
geometric-mean speedup of 6.25x, which we attribute to merging.

\mypara{Contributions}
In summary, we make the following contributions:
\begin{itemize}
\item \emph{Embeddings}, which make symbolic semantics configurable: users
replace the representation of types and the implementation of functions with
ordinary Haskell code over SMT primitives, on a per-property basis, and GHC
checks them through its coercion infrastructure.
\item \calc, a calculus for lazy, all-paths symbolic evaluation of System FC in
which values of any type can be merged symbolically, proved sound and
bounded-complete in Rocq, together with a formalisation of embeddings as a
translation based on coercion axioms (\cref{sec:embed}).
\item An implementation of \tool as a GHC plugin.
\item An evaluation on the \texttt{containers} and \texttt{iproute} libraries,
case studies for \texttt{IORef}, \texttt{Ptr}/\texttt{ForeignPtr} and \clash,
and a comparison with G2.
\end{itemize}

\section{Overview}
\label{sec:overview}
%
%
%
\sys is a GHC compiler plugin for compile-time property verification.
Properties are ordinary Haskell functions; \sys symbolically executes
them over all possible inputs, encodes the resulting path conditions as
SMT formulae, and asks Z3~\cite{z3} whether any input can falsify the property.
Much like QuickCheck~\cite{quickcheck}, \sys lets developers express properties
directly in Haskell and run them against inputs, but where QuickCheck
\emph{samples} inputs randomly, \sys \emph{exhausts} all paths and either
produces a proof of correctness or a guaranteed counterexample.

\mypara{Properties as Haskell Functions}
A property in \sys is a Haskell computation of type \lstinline{SymFC.Sym ()}.
Inside \lstinline{SymFC.Sym}, the call \lstinline{SymFC.fresh} creates a fresh
symbolic value of any type, and the call \lstinline{SymFC.valid} asks the SMT solver
whether a symbolic Boolean holds for every value of the symbolic inputs.
An annotation pragma marks the property for \sys and names the embeddings
that it is checked under.
As a first example, consider verifying that inserting a key into a
\lstinline{Data.Map} and immediately looking it up returns the inserted value:
\begin{lstlisting}
{-# ANN mapLookupInsert (SymFC [%'%embeddings]) #-}
mapLookupInsert :: SymFC.Sym ()
mapLookupInsert = do
  k <- SymFC.fresh @Int; v <- SymFC.fresh @Int
  let prop = Map.lookup k (Map.insert k v Map.empty) == Just v
  result <- SymFC.valid (SMT.boolean prop)
  case result of SymFC.None -> pure (); _ -> SymFC.halt
\end{lstlisting}
The annotation refers to the top-level binding \lstinline{embeddings} by name.
This binding holds the built-in embedding set, which provides symbolic
interpretations for GHC's numeric primitives.
The calls to \lstinline{SymFC.fresh} make \lstinline{k} and \lstinline{v} symbolic integers.
The rest of the body is ordinary Haskell: \lstinline{Map.insert}, \lstinline{Map.lookup}, and
\lstinline{(==)} are the real library functions from \lstinline{Data.Map.Strict}.
The \lstinline{SMT.boolean} function lifts the Haskell \lstinline{Bool} result into the
symbolic Boolean type \lstinline{SMT.Bool}.
\sys runs the property at compile time.
The call to \lstinline{SymFC.valid} returns \lstinline{SymFC.None} if no input can falsify the
property, and \lstinline{SymFC.Instance} if the solver finds an input that does.
The property calls \lstinline{SymFC.halt} in the second case, which stops compilation with an error.
Most properties follow this pattern, so we wrap it in a helper function:
\begin{lstlisting}
expectValid :: Bool -> SymFC.Sym ()
expectValid prop = do
  result <- SymFC.valid (SMT.boolean prop)
  case result of SymFC.None -> pure (); _ -> SymFC.halt
\end{lstlisting}

\mypara{Counterexample Generation}
\sys finds counterexamples for invalid properties.
Consider the following broken variant, which asserts membership
without the necessary precondition that the two keys are equal:
\begin{lstlisting}
{-# ANN mapMemberWrongKey (SymFC [%'%embeddings]) #-}
mapMemberWrongKey :: SymFC.Sym ()
mapMemberWrongKey = do
  k1 <- SymFC.fresh @Int; k2 <- SymFC.fresh @Int; v <- SymFC.fresh @Int
  expectValid $ Map.member k1 (Map.singleton k2 v)
\end{lstlisting}
The solver immediately finds an input that falsifies the property, for example
\lstinline{k1 = 0}, \lstinline{k2 = 1} and \lstinline{v = 0}.
Thus \lstinline{SymFC.valid} returns \lstinline{SymFC.Instance}, and compilation stops.

\mypara{Symbolic Algebraic Data Types}
\sys is not limited to primitive types.
It can symbolically execute code that operates over arbitrary algebraic
data types (ADTs), including recursive ones.
The \lstinline{Data.Map.Strict} and \lstinline{Data.Set} properties in the example above
illustrate this: both structures are internally balanced binary trees
(weight-balanced trees), implemented in roughly a thousand lines of Haskell.
\sys Symbolically executes the actual tree operations: \lstinline{insert},
\lstinline{delete}, rotations, rebalancing, exploring all execution paths through
the full implementation.
No specialized embeddings or abstract models are required for the containers;
\sys works directly on their source.

\mypara{Embeddings}
Symbolic execution of the full implementation is not always desirable.
For types implemented via foreign primitives or IO, the source code is
unavailable or too low-level to execute symbolically.
For types with rich SMT theories such as bitvectors or strings, a
direct abstract encoding is far more efficient than re-deriving the same
facts by executing thousands of lines of library code.
\sys addresses this with \emph{coercions as embeddings}: user-supplied mappings
from a Haskell type or function to an alternative interpretation.
For example, a common embedding is that of the base Haskell \lstinline{Integer} to the SMT solver primitive \lstinline{SMT.Integer}, which starts by specifying exactly this mapping, written \lstinline{Integer} \m{\mapsto} \lstinline{SMT.Integer}.
The mapping exposes a local instance of \lstinline{Embeddable}: a typeclass exported by \tool that witnesses that two types are equivalent at runtime.
We can use this local witness to give an embedding of the \lstinline{Integer} addition operation \lstinline{plusInteger}:
\begin{lstlisting}
plusEmbed :: Embeddable Integer SMT.Integer
          => Integer -> Integer -> Integer
plusEmbed = coerce SMT.addInteger
\end{lstlisting}
Here, we \lstinline{coerce} the \lstinline{SMT.Integer} addition to operate over values of type \lstinline{Integer}.
The instance of \lstinline{Embeddable} allows us to safely convert between values of type \lstinline{Integer} and \lstinline{SMT.Integer}.
With this, we can create a well-formed embedding \lstinline{plusInteger} \m{\mapsto} \lstinline{plusEmbed}.
\sys will provide the \lstinline{Embeddable} instance to \lstinline{plusEmbed}, after which its type matches up exactly to that of \lstinline{plusInteger}.

\mypara{Multiple Embeddings}
Most symbolic execution frameworks allow at most one embedding per type: at
worst only a handful of hardcoded built-in interpretations (e.g.\ G2~\cite{lazycounter}),
at best a single user-defined mapping to an SMT primitive (e.g.\
LiquidHaskell~\cite{liquidhaskell}).
\sys lifts this restriction by scoping embeddings per property.
Each property lists the embeddings it uses by name, and \sys combines them with the monoid operation:
\begin{lstlisting}
{-# ANN example (SymFC [%'%textAsSMTString, %'%int64AsBitVec64]) #-}
example :: SymFC.Sym ()
example = ...
\end{lstlisting}
This per-property scoping matters because there is often no single correct
embedding for a type.
Consider the \lstinline{Text} type from the \texttt{text} library.
A library \emph{user} verifying their own code wants \lstinline{Text}~$\mapsto$~\lstinline{SMT.String}:
SMT solvers provide a dedicated string theory that can reason about string
operations natively, and this is by far the most efficient encoding available.
Without the embedding, symbolic execution would have to unfold every string
operation into the low-level array allocations and pointer dereferences that
underlie \lstinline{Text}'s implementation, which quickly becomes intractable.
The library \emph{author} verifying the \texttt{text} implementation itself
wants the opposite: embedding \lstinline{Text} as an SMT string would hide
the very behavior they are trying to check.
Both are valid; \sys accommodates both by letting each property declare its
own embedding.

\mypara{Unrestricted Embeddings}
Another restriction commonly imposed on embeddings is that they can only map
directly to SMT types and functions.
Lifting this restriction opens up a wide variety of ways to model types which
would otherwise be unsupported.
A concrete example is the bitvector type in the \clash library~\cite{Clash}.
\lstinline{Clash.BitVec n} can range over any natural \lstinline{n}, including zero, whereas
\lstinline{SMT.BitVec n} requires \lstinline{n} to be positive, so a direct embedding is incomplete.
Instead, we can define \lstinline{Wrap.BitVec n}, which uses
\lstinline{SMT.BitVec n} for all positive sizes and a unit constructor for
zero.:
\begin{lstlisting}
data Wrap.BitVec n where
  BitVecZero :: Wrap.BitVec 0
  BitVecPos  :: 1 %$\leq$% n => SMT.BitVec n -> Wrap.BitVec n
\end{lstlisting}
\lstinline{BitVecZero} can only be instantiated when \lstinline{n} is zero;
\lstinline{BitVecPos} carries the evidence that \lstinline{1} $\leq$ \lstinline{n} alongside a
primitive SMT bitvector.
We can then embed \lstinline{Clash.BitVec n} as \lstinline{Wrap.BitVec n}, which is both
complete and uses the SMT solver primitive whenever the size is positive.

\mypara{Embedding IO}
The same principle lets us model IO in \tool.
Here \lstinline{RealWorld#}, the GHC token that sequences IO operations, is embedded as heap of type \lstinline{IntMap Any}: a map that may contain a value of any type, written \lstinline{RealWorld#} \m{\mapsto} \lstinline{IntMap Any}.
This heap can be indexed with a value of type \lstinline{Int}, resembling a pointer.
As such, the mutable reference type \lstinline{IORef} can be modeled by this pointer into our heap accordingly, written \lstinline{IORef a} \m{\mapsto} \lstinline{Int}.
With these type embeddings, we can embed the \lstinline{IORef} operations \lstinline{newIORef}, \lstinline{readIORef} and \lstinline{writeIORef} as a set of functions that manipulate this heap using its pointer.
No special solver support is needed to embed \lstinline{IO}: the model is written entirely in Haskell.
We will describe these embeddings in full detail in \cref{sec:case-study}.

\section{Theory}
\label{sec:theory}
We now present \calc, our core calculus for lazy, all-paths symbolic evaluation
based on \FC~\cite{systemfc}---the same calculus implemented in GHC.
Within \calc, we identify the sub-language \core: a standard instance of \FC
whose primitive types and operations match SMT supported theories.
\calc extends \core with symbolic variables and symbolic branches, and symbolic
evaluation is concrete evaluation extended with rules for these symbolic values
and for merging them.
Symbolic execution for Haskell such as G2~\cite{lazycounter, G2Q} explores
paths one at a time, splitting a symbolic value of an algebraic data type into
one path per constructor.
In \calc, a value of any type can instead be a single merged symbolic value, so
all paths are evaluated together.
All paths, merged values, and failed assertions and assumptions are
represented inside the resulting value.
We show that symbolic evaluation is sound and (bounded) complete with respect
to concrete evaluation of \core.
\Cref{sec:embed} then formalises embeddings as a translation on \core programs.

\subsection{Concrete Language}
\label{sec:theory:con:lang}
\Cref{fig:lang} defines \calc, where the sub-language \core consists of the
unshaded constructs.

\mypara{Terms}
\core is a lambda calculus and thus uses variables \m{x}, applications
\m{\app{e_f}{e_a}} and abstractions \m{\abs{x}{e}} as its basis of computation.

\mypara{Constructors and Literals}
\core supports data constructors \m{D} and their pattern match \m{\case{e}{\alts{a}}}.
In a pattern match, expression \m{e} is matched to any of the patterns in \m{\alts{a}} of the form \m{\alt{D}{\vect{x}}{e_r}}.
If any pattern matches, the right-hand side \m{e_r} is evaluated.
The arguments in \m{e} to data constructor \m{D} are bound to the variables
\m{\vect{x}} in the evaluation of \m{e_r}.
\core supports literals \m{l} and primitive operations \m{\prim} that mirror
those in SMT theories directly, for example, the literal \m{0x00} of type
\m{SMT.BitVec\ 8}, and the primitive bitvector addition operation \m{SMT.bvadd}.
%
%
%
%
%

\mypara{Types}
Abstractions \m{\abs{x}{e}} over a term variable are typed by function type \m{\tyInst_a \rightarrow \tyInst_r}.
Abstractions \m{\abs{x}{e}} over a type variable are typed by universal
quantification \m{\quant{x}{\tyInst}}, where \m{x} may occur free in
\m{\tyInst}.
A function type is a function constructor \m{(\rightarrow)} saturated through type applications \m{\app{\tyInst_f}{\tyInst_a}}.
A type constructor \m{T} may similarly be saturated through type application
\m{\app{\tyInst_f}{\tyInst_a}} which types any of its corresponding data
constructors \m{D}.
Like terms, types may be cast via \m{\cast{\tyInst}{\coInst}} which adjusts the
kind of the type.
As in System F---a subset of \FC---\core allows types \m{\tyInst} to appear in
its expression.
The inclusion of types in terms allows \core to support parametric polymorphism.
As type checking \core is standard, and equivalent to type checking \FC, we omit
it.

\begin{figure}[t]
\begin{minipage}[t]{0.53\linewidth}
\begin{tabular}[t]{rcll}
\multicolumn{4}{l}{\emphbf{Expression}} \\
\m{e}
  & $\bnf$   & \m{x} & \emph{variable} \\
  & $\spmid$ & \m{\abs{x}{e}} & \emph{abstraction} \\
  & $\spmid$ & \m{\app{e_f}{e_a}} & \emph{application} \\
  & $\spmid$ & \m{l} & \emph{literal} \\
  & $\spmid$ & \m{\prim} & \emph{primitive operation} \\
  & $\spmid$ & \m{D} & \emph{data constructor} \\
  & $\spmid$ & \m{\case{e}{\alts{a}}} & \emph{case split} \\
  & $\spmid$ & \m{\tyInst} & \emph{type} \\
  & $\spmid$ & \m{\cast{e}{\coInst}} & \emph{cast} \\
  & $\spmid$ & \m{\coInst} & \emph{coercion} \\
  & $\spmid$ & \m{b} & \emph{bottom} \\
\rowcolor{symrulebg}
\cellcolor{white}
  & $\spmid$ & \m{s} & \emph{symbolic variable} \\ [\jot]

\multicolumn{4}{l}{\emphbf{Bottom}} \\
\m{b}
  & $\bnf$   & \m{\app{raise}{e}} & \emph{error throw} \\
  & $\spmid$ & \m{\ub} & \emph{undefined behaviour} \\
\rowcolor{symrulebg}
\cellcolor{white}
  & $\spmid$ & \m{\unreachable} & \emph{unreachable} \\
\rowcolor{symrulebg}
\cellcolor{white}
  & $\spmid$ & \m{\budget} & \emph{bound exceeded} \\ [\jot]
\end{tabular}
\end{minipage}
\begin{minipage}[t]{0.45\linewidth}
\begin{tabular}[t]{rcll}
\multicolumn{4}{l}{\emphbf{Alternative}} \\
\m{a}
  & $\bnf$   & \m{\alt{D}{\vect{x}}{e}} & \emph{pattern} \\ [\jot]

\multicolumn{4}{l}{\emphbf{Type}} \\
\m{\tyInst}
  & $\bnf$   & \m{x} & \emph{variable} \\
  & $\spmid$ & \m{\quant{x}{\tyInst}} & \emph{quantification} \\
  & $\spmid$ & \m{\app{\tyInst_f}{\tyInst_a}} & \emph{application} \\
  & $\spmid$ & \m{(\rightarrow)} & \emph{function constructor} \\
  & $\spmid$ & \m{T} & \emph{type constructor} \\
  & $\spmid$ & \m{\cast{\tyInst}{\coInst}} & \emph{cast} \\
\multicolumn{4}{l}{\emphbf{Coercion}} \\
\m{\coInst}
  & $\bnf$   & \m{x} & \emph{variable} \\
  & $\spmid$ & \m{\quant{x}{\coInst}} & \emph{quantification} \\
  & $\spmid$ & \m{C} & \emph{axiom} \\
  & $\spmid$ & \m{\app{refl}{\tyInst}} & \emph{reflexive} \\
  & $\spmid$ & \m{\app{sym}{\coInst}} & \emph{symmetric} \\
  & $\spmid$ & \m{\coInst_l \circ \coInst_r} & \emph{transitive} \\
  & $\spmid$ & \m{\app{\coInst_f}{\coInst_a}} & \emph{application} \\
  & $\spmid$ & \m{\app{left}{\coInst}} & \emph{select left} \\
  & $\spmid$ & \m{\app{right}{\coInst}} & \emph{select right} \\
\end{tabular}
\end{minipage}
\caption{Syntax of \calc. \label{fig:lang}}
\end{figure}

\mypara{Casts and Coerions}
The defining feature of \FC compared to System F is the introduction of a type
safe cast \m{\cast{e}{\coInst}}, which uses a coercion \m{\coInst} to convert
the type of a term.
A coercion \m{\coInst : \tyInst_1 \sim \tyInst_2} is a witness of type equality
between types \m{\tyInst_1} and \m{\tyInst_2}.
We can use it to cast an expression \m{e} of type \m{\tyInst_1} to an expression
of type \m{\tyInst_2}, written \m{\cast{e}{\coInst}}.
Both the cast operation \m{\cast{e}{\coInst}} as well as coercions \m{\coInst}
are valid expressions in \core.
%
%
%
Haskell uses coercions, in a number of ways, most of which are opaque to the
user. We can, for example, define a data type \lstinline{Age} with the
\lstinline{newtype} keyword, as in \lstinline{newtype Age = MkAge SMT.Integer}.
In Haskell, this yields a new coercion axiom \m{C_{MkAge} : SMT.Integer \sim
Age}, which serves as a witness that \lstinline{SMT.Integer} and \lstinline{Age}
have the same representation at runtime.
When we construct an instance such as \lstinline{MkAge 0}, it is translated to
the cast \m{\cast{0}{C_{MkAge}}}.
Conversely, a pattern match \lstinline{case age of MkAge int -> ...} uses
\m{\cast{age}{\app{sym}{C_{MkAge}}}}, where \m{C_{MkAge} : Age \sim SMT.Integer}
makes use of the fact that coercions are symmetric.
%
%
%
%
%
%
%
%
Coercions allow us to construct a zero-cost alternative to wrapper data type
declarations, as coercions can be elided at runtime; they therefore form the
backbone of Haskell features like type families, generalized abstract data types
and functional dependencies. 
In \sys we use them to represent embeddings.

\mypara{Bottom}
A bottom value \m{b} short-circuits the evaluation of an expression.
\core supports bottom values through direct error calls via \m{\app{raise}{e}},
which halt evaluation.
We represent undefined behaviour with bottom value \m{\ub}; this may be
triggered, for example, by an incomplete pattern match.

\begin{figure}[t]
\begin{minipage}[t]{0.55\linewidth}
\begin{tabular}[t]{rcll}
\multicolumn{4}{l}{\emphbf{Value}} \\
\m{v}
  & $\bnf$   & \m{p} & \emph{primitive} \\
  & $\spmid$ & \m{\app{\prim}{\vect{t}}} & \emph{partial prim-op} \\
  & $\spmid$ & \m{\thunk{\subst}{\abs{x}{e}}} & \emph{closure} \\
  & $\spmid$ & \m{\app{D}{\vect{t}}} & \emph{data constructor} \\
  & $\spmid$ & \m{\tyInst} & \emph{type} \\
  & $\spmid$ & \m{\coInst} & \emph{coercion} \\
  & $\spmid$ & \m{\cast{v}{\coInst}} & \emph{cast} \\
  & $\spmid$ & \m{b} & \emph{bottom} \\
\rowcolor{symrulebg}
\cellcolor{white}
  & $\spmid$ & \m{\ife{p}{v_t}{v_f}} & \emph{symbolic branch} \\

\multicolumn{4}{l}{\emphbf{Thunk}} \\
\m{t}
  & $\bnf$   & \m{\thunk{\subst}{e}} & \emph{suspension} \\
\rowcolor{symrulebg}
\cellcolor{white}
  & $\spmid$ & \m{\ife{p}{t_t}{t_f}} & \emph{symbolic branch} \\
\end{tabular}
\end{minipage}
\begin{minipage}[t]{0.43\linewidth}
\begin{tabular}[t]{rcll}
\multicolumn{4}{l}{\emphbf{Primitive}} \\
\m{p}
  & $\bnf$   & \m{l} & \emph{literal} \\
\rowcolor{symrulebg}
\cellcolor{white}
  & $\spmid$ & \m{\app{\prim}{\vect{p}}} & \emph{saturated prim-op} \\
\rowcolor{symrulebg}
\cellcolor{white}
  & $\spmid$ & \m{s} & \emph{symbolic variable} \\

\multicolumn{4}{l}{\emphbf{Environment}} \\
\m{\subst}
  & $\bnf$   & \m{\{ x \mapsto t \}} & \emph{substitution map} \\ [\jot]

\rowcolor{symrulebg}
\multicolumn{4}{l}{\emphbf{Path Condition}} \\
\rowcolor{symrulebg}
\m{\pathc}
  & $\bnf$   & \m{p} & \emph{conditional} \\ [\jot]
\end{tabular}
\end{minipage}
\caption{Weak-Head Normal Form Values of \calc. \label{fig:values}}
\end{figure}

\subsection{Concrete Evaluation}
\label{sec:theory:con:eval}
Lazy evaluation places an expression into weak head normal form
(WHNF)~\cite{peyton-jones-1987}.
We define WHNF value \m{v} in \Cref{fig:values}; grayed out parts of the Figure
concern symbolic evaluation and can be ignored for now.
Most constructs are shared directly between expressions and values, however, as
the head of a value must be evaluated, values cannot be a variables or a pattern
matches.

\mypara{Thunk}
Lazy evaluation keeps track of thunks: a thunk \m{t} represents a suspended
computation that may be forced when its value is required.
A thunk consists of an expression together with an environment \m{\subst} in
which the expression should be evaluated, where environments map variables to
their respective thunk.
An abstraction cannot be evaluated any further: it would first need to be
applied before we can perform beta reduction.
As such, closure \m{\thunk{\subst}{\abs{x}{e}}} is a value, which is itself a
special case of a thunk.
Thunks also appear as the arguments to data constructors and partially applied primitive operations.
These arguments are to be evaluated under the environment they were defined in,
which is why they are stored as thunks.
Application thus does not appear by itself in a value: data constructors and partially applied primitive operations track their arguments directly.
A primitive operation can never appear saturated as this would be evaluated to a literal.

\mypara{Rules}
Next, we define a lazy, big-step evaluation semantics for \core with the
judgement \m{\constep{\subst}{e}{v}}, which says that expression \m{e}
evaluates to value \m{v} under environment \m{\subst}.
We present the evaluation rules of \core in \Cref{fig:red-rules}.
The rules follow \FC exactly.
The rules \m{[Lit]}, \m{[Con]}, \m{[Prim]}, \m{[Bot]}, \m{[Coercion]} and
\m{[Type]} evaluate an expression directly to their respective value equivalent.
We write $\tyInst[\subst]$ and $\coInst[\subst]$ for capture-avoiding
instantiation of the free type and coercion variables, using the bindings in
$\subst$.
\m{[Lam]} attaches the current environment to the abstraction to form a closure.
We define a number of helper judgments for application, case splits and thunks in \Cref{fig:helper}.
Judgement \m{\vdash \app{v_f}{t} \Downarrow_{apply} v_r} describes how to apply
a thunk to a value.
For example, \m{[App{-}Clos]} applies a thunk to a closure by extending the
environment with a mapping from the binder to the thunk.
Judgement \m{\subst; \vect{a} \vdash v_s \Downarrow_{case} v_r} performs a
pattern match on a value given patterns~$\vect{a}$.
For example, \m{[Case{-}Con]} allows us select the corresponding data constructor if it exists.
Judgement \m{\subst \vdash t \Downarrow_{thunk} v} allows us to force a thunk.
In \core, only rule \m{[Thunk{-}Susp]} is supported.

\mypara{Example}
We illustrate the evaluation of \core with the toy example in \Cref{fig:eval-example}. The derivation tree
grows downwards and the axioms required by a rule are indented to the right. We
start the evaluation under an empty environment \m{\emptyset}.

\begin{figure}[t]
\begin{minipage}{\linewidth}
\dirtree{%
.1
  {\m
    {\constep
      {\emptyset}
      {\app
        {(\abs{x}{\case{x}{Unit \rightarrow 0}})}
        {\ub}
      }
      {\ub}
    }
    \hfill
    \m{[App]}
  }.
.2
  {\m
    {\constep
      {\emptyset}
      {\abs{x}{\case{x}{Unit \rightarrow 0}}}
      {\thunk{\emptyset}{\abs{x}{\case{x}{Unit \rightarrow 0}}}}
    }
    \hfill
    \m{[Lam]}
  }.
.2
  {\m
    { \vdash
      {\thunk{\emptyset}{\abs{x}{\case{x}{Unit \rightarrow 0}}}\ \thunk{\emptyset}{\ub}}
      \Downarrow_{apply}
      {\ub}
    }
    \hfill
    \m{[App{-}Clos]}
  }.
.3
  {\m
    {\constep
      {\{ x \rightarrow \thunk{\emptyset}{\ub} \}}
      {\case{x}{Unit \rightarrow 0}}
      {\ub}
    }
    \hfill
    \m{[Case]}
  }.
.4
  {\m
    {\constep
      {\{ x \rightarrow \thunk{\emptyset}{\ub} \}}
      {x}
      {\ub}
    }
    \hfill
    \m{[Var]}
  }.
.5
  {\m
    { \vdash
      {\thunk{\emptyset}{\ub}}
      \Downarrow_{thunk}
      {\ub}
    }
    \hfill
    \m{[Thunk{-}Susp]}
  }.
.6
  {\m
    {\constep
      {\emptyset}
      {\ub}
      {\ub}
    }
    \hfill
    \m{[Bot]}
  }.
.4
  {\m
    { \{ x \rightarrow \thunk{\emptyset}{\ub} \}
      ; Unit \rightarrow 0
      \vdash
      {\ub}
      \Downarrow_{case}
      {\ub}
    }
    \hfill
    \m{[Case{-}Bot]}
  }.
}
\end{minipage}
\caption{Evaluation of a \core example. \label{fig:eval-example}}
\end{figure}
Rule \m{[App]} has two sub-goals: evaluate the function expression and apply the resulting value to the argument thunk.
The function is an abstraction so \m{[Lam]} allows us to evaluate it to a closure using the current (empty) environment.
We then apply the argument as a thunk to this closure through \m{[App{-}Clos]}.
This places the thunk in the environment under the binder \m{x} for evaluation of the case split.
\m{[Case]} requires us to first evaluate the scrutinee \m{x} first.
We lookup the variable with \m{[Var]} and force the resulting thunk with \m{[Thunk{-}Susp]}.
Finally, we end up with \m{\ub} through \m{[Bot]}.
As per \m{[Case]}, we need to pattern match this resulting bottom value.
\m{[Case{-}Bot]} allows us to discharge this branch, terminating the evaluation with \m{\ub} as its result.

\begin{figure}
\centering
\begin{minipage}{0.45\linewidth}
\begin{prooftree}
\axiomc{\subst (x)=t}
\axiomc{\forceb{\pathc}{n-1}{t}{v}}
\rightlabel{Var}
\binaryc
  {\bigstepb
    {\pathc}
    {\subst}
    {n}
    {x}
    {v}
  }
\end{prooftree}
\end{minipage}
\begin{minipage}{0.25\linewidth}
\begin{prooftree}
\axiomc{}
\rightlabel{Lit}
\unaryc{\bigstep{\pathc}{\subst}{l}{l}}
\end{prooftree}
\end{minipage}
\begin{minipage}{0.28\linewidth}
\begin{symprooftree}
\axiomc{}
\rightlabel{Sym{-}Var}
\unaryc{\bigstep{\pathc}{\subst}{s}{s}}
\end{symprooftree}
\end{minipage}

\vspace{5pt}

\begin{minipage}{0.59\linewidth}
\begin{prooftree}
\axiomc{\bigstepb{\pathc}{\subst}{n-1}{e_f}{v_f}}
\axiomc{\applyb{\pathc}{n-1}{v_f}{\thunk{\subst}{e_a}}{v_r}}
\rightlabel{App}
\binaryc
  {\bigstepb
    {\pathc}
    {\subst}
    {n}
    {\app{e_f}{e_a}}
    {v_r}
  }
\end{prooftree}
\end{minipage}
\hfill
\begin{minipage}{0.40\linewidth}
\begin{symprooftree}
\axiomc
  {\neg SAT(\pathc)
  }
\rightlabel{Prune}
\unaryc
  {\bigstep
    {\pathc}
    {\subst}
    {e}
    {\unreachable}
  }
\end{symprooftree}
\end{minipage}

\vspace{5pt}

\begin{minipage}{0.43\linewidth}
\begin{prooftree}
\axiomc
  {\bigstep
    {\pathc}
    {\subst}
    {e}
    {v}
  }
\axiomc
  {v' = cast(v, \coInst)
  }
\rightlabel{Cast}
\binaryc
  {\bigstep
    {\pathc}
    {\subst}
    {\cast{e}{\coInst}}
    {v'}
  }
\end{prooftree}
\end{minipage}
\begin{minipage}{0.37\linewidth}
\begin{prooftree}
\axiomc{}
\rightlabel{Lam}
\unaryc
  {\bigstep
    {\pathc}
    {\subst}
    {\abs{x}{e}}
    {(\subst, \abs{x}{e})}
  }
\end{prooftree}
\end{minipage}
\begin{minipage}{0.18\linewidth}
\begin{prooftree}
\axiomc{}
\rightlabel{Bot}
\unaryc
  {\bigstep
    {\pathc}
    {\subst}
    {b}
    {b}
  }
\end{prooftree}
\end{minipage}

\vspace{5pt}

\begin{minipage}{0.60\linewidth}
\begin{prooftree}
\axiomc
  {\bigstepb
    {\pathc}
    {\subst}
    {n-1}
    {e_s}
    {v_s}
  }
\axiomc{\foldb{\pathc}{\subst}{n-1}{v_s}{\vect{a}}{v_r}}
\rightlabel{Case}
\binaryc
  {\bigstepb
    {\pathc}
    {\subst}
    {n}
    {\case{e_s}{\alts{a}}}
    {v_r}
  }
\end{prooftree}
\end{minipage}
\hfill
\begin{minipage}{0.37\linewidth}
\begin{symprooftree}
\axiomc{}
\rightlabel{Bound}
\unaryc
  {\bigstepb
    {\pathc}
    {\subst}
    {0}
    {e}
    {\budget}
  }
\end{symprooftree}
\end{minipage}

\vspace{5pt}

\begin{minipage}{0.26\linewidth}
\begin{prooftree}
\axiomc{\coInst' = \coInst[\subst]}
\rightlabel{Coercion}
\unaryc
  {\bigstep
    {\pathc}
    {\subst}
    {\coInst}
    {\coInst'}
  }
\end{prooftree}
\end{minipage}
\hfill
\begin{minipage}{0.25\linewidth}
\begin{prooftree}
\axiomc{\tyInst' = \tyInst[\subst]}
\rightlabel{Type}
\unaryc
  {\bigstep
    {\pathc}
    {\subst}
    {\tyInst}
    {\tyInst'}
  }
\end{prooftree}
\end{minipage}
\hfill
\begin{minipage}{0.19\linewidth}
\begin{prooftree}
\axiomc{}
\rightlabel{Con}
\unaryc{\bigstep{\pathc}{\subst}{D}{D}}
\end{prooftree}
\end{minipage}
\hfill
\begin{minipage}{0.22\linewidth}
\begin{prooftree}
\axiomc{}
\rightlabel{Prim}
\unaryc{\bigstepb{\pathc}{\subst}{n}{\prim}{\prim}}
\end{prooftree}
\end{minipage}

\caption{Evaluation rules of \calc (rules with a gray background are symbolic rules that do not apply to concrete evaluation). Where the bound is omitted, a rule concludes at a
bound \m{n \ge 1} and its premises are at \m{n-1}, with \m{\infty - 1 = \infty}; at bound
\m{0} only \m{Bound} applies. \m{\epsilon} is the empty sequence of thunks, and
\m{D} is the constructor value \m{\app{D}{\vect{t}}} with no thunks. \label{fig:red-rules}}
\end{figure}

\begin{figure}[t]
\centering
\noindent\hfill$\displaystyle
\inferrule*[right=\m{App{-}Con}]
  { }
  {\m{\applyb{\pathc}{n}{(\app{D}{\vect{t_a}})}{t}{\app{D}{(\vect{t_a}\ t)}}}}$
\qquad
$\displaystyle
\inferrule*[right=\m{App{-}Par{-}Prim}]
  {\m{|\vect{t_a}\ t| < arity(\prim)}}
  {\m{\applyb
    {\pathc}
    {n}
    {(\app{\prim}{\vect{t_a}})}
    {t}
    {\app{\prim}{(\vect{t_a}\ t)}}}}$
\hfill\mbox{}\par
\vspace{6pt}
\noindent\hfill$\displaystyle
\inferrule*[right=\m{App{-}Sat{-}Prim}]
  {\m{|\vect{t_a}\ t| = arity(\prim)}}
  {\m{\applyb
    {\pathc}
    {n}
    {(\app{\prim}{\vect{t_a}})}
    {t}
    {reduce{-}prim(\app{\prim}{(\vect{t_a}\ t)})}}}$
\qquad
$\displaystyle
\inferrule*[right=\m{App{-}Bot}]
  { }
  {\m{\applyb{\pathc}{n}{b}{t}{b}}}$
\hfill\mbox{}\par
\vspace{6pt}
\noindent\hfill$\displaystyle
\inferrule*[right=\m{App{-}Cast}]
  { \m{\coInst : (\tyInst_a \rightarrow \tyInst_r) \sim (\tyInst_a' \rightarrow \tyInst_r')}
  \\ \m{\coInst_a = sym\ (right\ (left\ \coInst))}
  \\ \m{\applyb{\pathc}{n}{v}{(\cast{t}{\coInst_a})}{v_r}}}
  {\m{\applyb{\pathc}{n}{(\cast{v}{\coInst})}{t}{\cast{v_r}{right\ \coInst}}}}$
\hfill\mbox{}\par
\vspace{6pt}
\noindent\hfill$\displaystyle
\inferrule*[right=\m{App{-}Clos}]
  {\m{\bigstepb{\pathc}{\subst' \{ x \mapsto t\}}{n}{e_b}{v_r}}}
  {\m{\applyb{\pathc}{n}{\thunk{\subst'}{\abs{x}{e_b}}}{t}{v_r}}}$
\qquad
$\displaystyle
\inferrule*[right=\m{Case{-}Otherwise}]
  {\text{no alternative of } \m{\vect{a}} \text{ matches } \m{v}}
  {\m{\foldb{\pathc}{\subst}{n}{v}{\vect{a}}{\ub}}}$
\hfill\mbox{}\par
\vspace{6pt}
\noindent\hfill$\displaystyle
\symrule{\inferrule*[right=\m{App{-}If}]
  {\m{\applyb{\pathc \land p}{n}{v_t}{t}{v_t'}} \\ \m{\applyb{\pathc \land \neg p}{n}{v_f}{t}{v_f'}}}
  {\m{\applyb{\pathc}{n}{(\ife{p}{v_t}{v_f})}{t}{ite(p, v_t', v_f')}}}}$
\qquad
$\displaystyle
\inferrule*[right=\m{Case{-}Bot}]
  { }
  {\m{\foldb{\pathc}{\subst}{n}{b}{\vect{a}}{b}}}$
\hfill\mbox{}\par
\vspace{6pt}
\noindent\hfill$\displaystyle
\inferrule*[right=\m{Thunk{-}Susp}]
  {\m{\bigstepb{\pathc}{\subst}{n}{e}{v}}}
  {\m{\forceb{\pathc}{n}{\thunk{\subst}{e}}{v}}}$
\qquad
$\displaystyle
\symrule{\inferrule*[right=\m{Thunk{-}If}]
  {\m{\forceb{\pathc \land p}{n}{t_t}{v_t}} \\ \m{\forceb{\pathc \land \neg p}{n}{t_f}{v_f}}}
  {\m{\forceb{\pathc}{n}{\ife{p}{t_t}{t_f}}{ite(p, v_t, v_f)}}}}$
\hfill\mbox{}\par
\vspace{6pt}
\noindent\hfill$\displaystyle
\inferrule*[right=\m{Case{-}Con}]
  {\m{find(D, \vect{a}) = \alt{D}{\vect{x}}{e_r}} \\
   \m{\bigstepb{\pathc}{\subst\{\vect{x \mapsto t}\}}{n}{e_r}{v_r}}}
  {\m{\foldb{\pathc}{\subst}{n}{\app{D}{\vect{t}}}{\vect{a}}{v_r}}}$
\hfill\mbox{}\par
\vspace{6pt}
\noindent\hfill$\displaystyle
\symrule{\inferrule*[right=\m{Case{-}If}]
  {\m{\foldb{\pathc \land p}{\subst}{n}{v_t}{\vect{a}}{v_t'}} \\
   \m{\foldb{\pathc \land \neg p}{\subst}{n}{v_f}{\vect{a}}{v_f'}}}
  {\m{\foldb{\pathc}{\subst}{n}{\ife{p}{v_t}{v_f}}{\vect{a}}{ite(p, v_t', v_f')}}}}$
\hfill\mbox{}\par
\caption{Applying a value to a thunk argument, forcing a thunk and pattern matching a value. No alternative matches \m{v} when \m{v \equiv \app{D}{\vect{t}}} and \m{find(D, \vect{a})} is undefined, or when \m{v} is neither a constructor, a bottom nor a symbolic branch. \label{fig:helper}}
\end{figure}

\subsection{Symbolic Language}
\label{sec:theory:sym:lang}
\core is a standard instance of \FC where we adjusted its presentation to
closely resemble the typed intermediate representation used by GHC.
We now extend \core to our full calculus \calc, which supports symbolic
evaluation.
The extension is given by the \colorbox{symrulebg}{grayed} definitions in
\Cref{fig:lang} and \Cref{fig:values}.

\mypara{Symbolic Variable}
\calc supports symbolic variables \m{s}, which an SMT solver can reason about.
Their inclusion drives all other additions to the calculus and evaluation rules.
Symbolic variables are free in an expression: they are never resolved from the
environment \m{\subst} and occur directly in both expressions \m{e} and in
values \m{v}, grouped under primitives \m{p} in the latter.

\mypara{Primitive}
Unlike concrete evaluation, we cannot always evaluate a saturated primitive operation to a literal.
For example, there is no way to reduce \m{\app{\app{SMT.bvadd}{s_l}}{s_r}} to a literal as both arguments are symbolic.
We thus allow for saturated primitive operations, which we group with literals
and symbolic variables under primitive \m{p}.
Primitives are directly supported by an SMT solver.

\mypara{Symbolic Branching}
Symbolic variables can only have primitive types supported by the solver.
We can define a non-primitive expression symbolically through the use of the
primitive branching operation \m{ite :: SMT.Bool \rightarrow a \rightarrow a
\rightarrow a} over a symbolic conditional.
For example, we can define a symbolic instance for an expression of type
\m{\app{Maybe}{(\app{SMT.BitVec}{8})}} as
\m{\app{\app{\app{ite}{s_b}}{Nothing}}{(\app{Just}{s_{bv}})}}.

\noindent
A conditional may remain symbolic during evaluation and thus the branch needs to be kept explicitly.
We extend values with an explicit symbolic branch \m{\ife{p}{v_t}{v_f}}.
The symbolic branch follows the ordered guards representation introduced by Grisette \cite{Grisette}.
This representation permits us to merge common sub-values in leaves of the tree during evaluation.
For example, the value \m{\ife{p}{\app{D}{t_1}}{\app{D}{t_2}}} can be merged
into the value \m{\app{D}{(\ife{p}{t_1}{t_2})}} as the leaves share the same
data constructor \m{D}.
To accomodate merging, we permit symbolic branches in thunks.
We do however not merge thunks directly.

\mypara{Bottom}
We introduce two new bottom values that can only occur during symbolic
evaluation.
\begin{enumerate}[leftmargin=*]
\item
\emph{Unreachable} \m{\unreachable} represents a path that cannot be reached.
There is no reason to continue evaluating an unreachable path, hence, this
bottom value allows us to halt evaluation.

\item
\emph{Bound} \m{\budget} allows us to halt evaluation early when we exceed the evaluation depth.
We can use \m{\budget} to find a symbolic instance for expressions that contain
a non-terminating branch by representing non-termination via \m{\budget}.
This is semantically different from pruning with unreachable \m{\unreachable} as
the branch remains reachable.
Instead, we ignore the infinite path to allow the symbolic evaluation to
terminate in order to inspect any finite branches.
\end{enumerate}

\noindent
The bottom values \m{\unreachable} and \m{\ub} respectively allow us to
naturally define assumptions and assertions within expressions.
To assume predicate \m{p} holds in the evaluation of \m{e}, we can introduce a
conditional with \m{\unreachable} in its false branch: \m{\app{ite}{p\ e\
\unreachable}}. Similarly, we can assert that \m{p} holds in the evaluation of
\m{e} using expression \m{\app{ite}{p\ e\ \ub}}.
As a consequence, the evaluation semantics of \calc do not require a global
store of assertions and assumptions; these are directly represented inside
symbolic expression.

\subsection{Symbolic Evaluation}
\label{sec:theory:sym:eval}
We extend the concrete big-step evaluation semantic of \core with a path condition \m{\pathc} and evaluation depth \m{n}.
The path condition \m{\pathc} is a Boolean primitive \m{p} used to prune unreachable expressions through \m{\unreachable}.
The evaluation depth \m{n} tracks when to halt a branch through a bound \m{\budget}.
The final judgement is of the form $\m{\bigstepb{\pathc}{\subst}{n}{e}{v}}$.
It evaluates expression \m{e} to value \m{v} under path condition \m{\pathc} and
derivation depth bound \m{n \in \nat_\infty}.
We omit the index when \m{n = \infty}.
The judgements that enable \emph{symbolic} evaluation are \colorbox{symrulebg}{grayed} in \Cref{fig:red-rules}.
\m{[Sym{-}Var]} returns symbolic variables, as is.
\m{[Prune]} abandons an infeasible path with \m{\unreachable}.
Reachability is concluded through satisfiability check \m{SAT}.
\m{[Bound]} enforces the evaluation depth bound with \m{\budget} if the depth \m{n} is 0.

\mypara{Case Split}
\m{[Case]} first evaluates the scrutinee \m{e_s} to value \m{v_s} before pattern matching on it within judgement \m{\Downarrow_{case}}.
If value \m{v_s} is itsef a symbolic branch, then we pattern match both leaves individually with \m{[Case{-}If]}.
Any other value is handled as in concrete execution.
Without merging, this pattern match would be slow: each symbolic branch would be
evaluated individually, which yields path enumeration.
Instead, we want to merge common prefixes before pattern matching.
%

\begin{figure}[t]
\centering
\[
\m{ite2(p, v_t, v_f) = \begin{cases}
\m{\app{D}{\vect{(\ite{p}{t_1}{t_2})}}}
  & \m{v_t \equiv \app{D}{\vect{t_1}} \land v_f \equiv \app{D}{\vect{t_2}} \land |\vect{t_1}| = |\vect{t_2}|}
  \\
\m{\app{\prim}{\vect{(\ite{p}{t_1}{t_2})}}}
  & \m{v_t \equiv \app{\prim}{\vect{t_1}} \land v_f \equiv \app{\prim}{\vect{t_2}} \land |\vect{t_1}| = |\vect{t_2}|}
  \\
\m{\app{ite}{p \ p_t \ p_f}}
  & \m{v_t \equiv p_t} \land \m{v_f \equiv p_f}
  \\
\m{\cast{ite(p, v_t', v_f')}{\coInst}}
  & \m{v_t \equiv \cast{v_t'}{\coInst} \land v_f \equiv \cast{v_f'}{\coInst}}
  \\
\m{b}
  & \m{v_t \equiv b \land v_f \equiv b}
  \\
\m{\tyInst}
  & \m{v_t \equiv \tyInst \land v_f \equiv \tyInst}
  \\
\m{\coInst_t}
  & \m{v_t \equiv \coInst_t : \tyInst_1 \sim \tyInst_2} \land \m{v_f \equiv \coInst_f : \tyInst_1 \sim \tyInst_2}
  \\
\m{\ife{p}{v_t}{v_f}}
  & \m{otherwise}
  \\
\end{cases}}
\]
\caption{Merging two leaves of a branch. \label{fig:merge}}
\end{figure}

\mypara{Merge}
We follow the merging algorithm of ordered guards, as defined in
\cite{Grisette}.
\Cref{fig:merge} shows how to merge common prefixes of two values with function
\m{ite2}, where we use $\equiv$ for syntactic equality.
A matching data constructor between branches allows us to push the conditional statement into its arguments.
The arguments are thunks and remain unmerged until they are forced.
Partial applications of the same primitive merge in the same way.
Primitives merge into the SMT term \m{\app{ite}{p\ p_t\ p_f}}, which the SMT theory may simplify.
Casts merge only if their coercions witness the same equality.
We arbitrarily pick one of the coercions for the resulting cast and merge their
bodies.
Similarly, coercions can be unified by arbitrarily picking one, if they witness
the same type equality.
Concrete types and bottom values are only merged if they are exactly equal.
If no prefix of the expression matches, then we simply retain the expression as a symbolic branch.

\mypara{Merge Invariant}
The full function \m{ite(p, v_t, v_f)}---whose definition we elide---follows the
same idea but merges values for any branch depth.
We achieve this by grouping common values through a standard merge sort before
attempting to merge them.
We can both trivially and arbitrarily pick a total ordering for values.
We never manually construct a symbolic branch for values: their construction
always goes through \m{ite(p, v_t, v_f)}, which ensures that values are always
merged.

\mypara{Application and Thunks}
The final extension to the evaluation rules is for application and forcing.
\m{[App{-}If]} applies a thunk to a symbolic branch by distributing it along both leaves.
We attempt to merge the resulting leaves through a call to \m{ite(p, v_t, v_f)}.
When we force a symbolic branch in a thunk with \m{Thunk{-}If}, we similarly merge the resulting values.

\mypara{Example}
We modify our previous example to run on a symbolic argument that
non-deterministically represents \m{Unit} or \m{\ub}.
The starting path condition is \m{true}, and the depth bound is \m{\infty}.
This gives us the following judgement:
\[
\m{\bigstep
  {true}
  {\emptyset}
  {\app
    {(\abs{x}{\case{x}{Unit \rightarrow 0}})}
    {(ite\ s_c\ Unit\ \ub)}
  }
  {\ife{s_c}{0}{\ub}}
}
\]
The evaluation follows the concrete example.
The most interesting step is the case split \m{ite\ s_c\ Unit\ \ub} after
forcing it within \m{[Case]}.
\[
\m{
  true
  ;
  \subst
  ; Unit \rightarrow 0
  \vdash
  \ife{s_c}{Unit}{\ub}
  \Downarrow_{case}
  \ife{s_c}{0}{\ub}
}
\]
We use \m{[Case{-}If]} to distribute the pattern match along both symbolic branches.
This way, we can individually pattern match the \m{Unit} data constructor with \m{[Case{-}Con]} and the bottom value \m{\ub} with \m{[Case{-}Bot]}.
We take every branch of an alternative at most once because common data constructors are merged.

\subsection{Soundness and Completeness}
\label{sec:theory:sound}
Next, we prove soundness and completeness of \calc with respect to \core.
\begin{theorem}[Concrete determinism]
\label{thm:concrete-determinism}
Let \m{e} be a \core program in a concrete environment \m{\subst}.
If \m{\constep{\subst}{e}{v_1}} and \m{\constep{\subst}{e}{v_2}}, then \m{v_1 = v_2}.
\end{theorem}

\noindent
\cref{thm:concrete-determinism} says that any \core program that terminates
evaluates to only one value.

\mypara{Instances}
A symbolic run is only useful if it says something about concrete runs, so we
need to define which concrete runs a symbolic run represents.
Let a \emph{model} \m{\sigma} map each symbolic variable of a run to a literal; this is what
the solver returns when a path condition is satisfiable, and we write
\m{\sigma \models \pathc} when \m{\sigma} satisfies \m{\pathc}.
We write \m{\sigma(p)} for the literal that the SMT theory gives the primitive value
\m{p} under \m{\sigma}.
We write \m{\instance{e_s}{\sigma}{e_c}} when the \core expression \m{e_c} is an instance of
the \calc expression \m{e_s} under \m{\sigma}, and likewise for values, thunks and
environments.
\begin{figure}[t]
\centering
\begin{mathpar}
\inferrule*[right=\m{I{-}IfT}]
  {\m{\sigma \models p} \\ \m{\instance{v_t}{\sigma}{v}}}
  {\m{\instance{\ife{p}{v_t}{v_f}}{\sigma}{v}}}
\and
\inferrule*[right=\m{I{-}IfF}]
  {\m{\sigma \models \neg p} \\ \m{\instance{v_f}{\sigma}{v}}}
  {\m{\instance{\ife{p}{v_t}{v_f}}{\sigma}{v}}}
\\
\inferrule*[right=\m{I{-}Sym}]
  { }
  {\m{\instance{s}{\sigma}{\sigma(s)}}}
\and
\inferrule*[right=\m{I{-}Denote}]
  {\m{|\vect{p}| = arity(\prim)} \\ \m{\sigma(\app{\prim}{\vect{p}}) = l}}
  {\m{\instance{\app{\prim}{\vect{p}}}{\sigma}{l}}}
\end{mathpar}
\caption{Key rules of the instance relation (congruence rules omitted; full relation in \Cref{fig:instances}). \label{fig:instances-key}}
\end{figure}
The relation matches the two sides syntactically constructor by constructor with additional non-trivial rules shown in \Cref{fig:instances-key}:
\m{\instance{s}{\sigma}{\sigma(s)}} for a symbolic variable \m{s} (\m{I{-}Sym});
\m{\instance{\app{\prim}{\vect{p}}}{\sigma}{l}} when the primitive value
\m{\app{\prim}{\vect{p}}} is saturated and has the value \m{l} under \m{\sigma}, so SMT
terms need not match syntactically (\m{I{-}Denote});
and \m{\instance{\ife{p}{v_t}{v_f}}{\sigma}{v}} when \m{\sigma} satisfies the branch
condition \m{p} and \m{\instance{v_t}{\sigma}{v}} (\m{I{-}IfT}), with \m{v_f} instead when
\m{\sigma} falsifies \m{p} (\m{I{-}IfF}).
Branches of thunks have the same two rules.

\noindent
For the rest of this section let \m{e_c} be a \core program in a concrete environment
\m{\subst_c}, let \m{e_s} be a \calc program in \m{\subst_s}, and let
\m{\sigma \models \pathc} with \m{\instance{e_s}{\sigma}{e_c}} and
\m{\instance{\subst_s}{\sigma}{\subst_c}}.

\begin{theorem}[Soundness]
\label{thm:soundness}
If \m{\bigstep{\pathc}{\subst_s}{e_s}{v_s}}, then \m{\constep{\subst_c}{e_c}{v_c}} for some
\m{v_c} with \m{\instance{v_s}{\sigma}{v_c}}.
\end{theorem}

\noindent
\cref{thm:soundness} rules out spurious counterexamples.
When \tool finds a model \m{\sigma} for a symbolic program \m{e_s}, soundness
says that instantiating \m{e_s} under \m{\sigma} and running the result
concretely really does produce the failure the symbolic run found.
Completeness needs one more condition on \m{e_s}.
\m{e_s} is \emph{budget-total} when there is an \m{h} such that for every \m{n \ge h}
there is a \m{v} with \m{\bigstepb{\pathc}{\subst_s}{n}{e_s}{v}}.
This excludes programs that get stuck on a feasible path, even one the model does not select.

\begin{theorem}[Bounded completeness]
\label{thm:completeness}
Let \m{e_s} be budget-total.
If \m{\constep{\subst_c}{e_c}{v_c}}, then \m{\exists h. \forall n \in \nat.\,n \ge h}
there is a \m{v_s} with \m{\bigstepb{\pathc}{\subst_s}{n}{e_s}{v_s}} and
\m{\instance{v_s}{\sigma}{v_c}}.
\end{theorem}

\noindent
Budget-totality is needed only for the existence of \m{v_s}.
Without it, the same \m{h} still gives \m{\instance{v_s}{\sigma}{v_c}} for every
\m{n \ge h} and every \m{v_s} with \m{\bigstepb{\pathc}{\subst_s}{n}{e_s}{v_s}}.
\cref{thm:completeness} rules out missed failures.
If some input makes the program fail concretely, then running \m{e_s} at any
finite bound above \m{h} produces a symbolic value covering that concrete
failure.
The bound is needed because the reductions of \m{\Downarrow^n_{case}}, \m{\Downarrow^n_{apply}} and \m{\Downarrow^n_{thunk}} demand a value from both leaves of a branch.
A concrete run follows one path, but the symbolic run has to produce a value for every leaf,
and a leaf the model does not select may loop; with no bound the whole run then has no
value at all, even though the concrete run terminates.
Under a bound, the non-terminating leaf becomes \m{\budget} instead, and the run
still delivers a value on the path the model does take.
%
%
The theorems are proved under assumptions on what we treat as opaque: the
solver, substitution of types and coercions, \m{reduce{-}prim} and \m{cast}. We
assume, for example, that the solver accepts \m{true}, and that it accepts every
path condition that some model satisfies.
%
We mechanised the three theorems in Rocq, with no axioms; \Cref{app:proofs}
gives the proof sketches.

\section{Embeddings}
\label{sec:embed}
An embedding enables us to replace the definition of a type or function with an
alternative interpretation.
For symbolic evaluation, the replacement is usually one the solver understands, such as \m{Integer \mapsto SMT.Integer}.
Embeddings do not depend on symbolic evaluation, though: they are a general way to give types and terms another representation.
An embedding has two parts.
A \emph{type embedding} says how a type is represented, and a \emph{term embedding} says how a function is implemented.
The two must agree, and this section shows how we make them agree.

\mypara{Type Embeddings}
A type embedding \m{\tyInst_{orig} \mapsto \tyInst_{interp}} declares a coercion axiom \m{C : \tyInst_{orig} \sim \tyInst_{interp}}.
The axiom is a witness that the original type and its interpretation are equal, so a cast can move a value from one to the other (\Cref{sec:theory:con:lang}).
For example, \m{Integer \mapsto SMT.Integer} declares \m{C_{Integer} : Integer \sim SMT.Integer}.
The original may have parameters: \m{[\alpha] \mapsto \app{SMT.Seq}{\alpha}} declares the axiom \m{C_{[]} : \quant{\alpha}{[\alpha] \sim \app{SMT.Seq}{\alpha}}}.
Both types must have the same kind---the type of a type.

\mypara{Term Embeddings}
A term embedding \m{x \mapsto e} replaces the definition of \m{x} with the interpretation \m{e}.
The interpretation must have the type of the original, or the program around it no longer type checks.
The embedding \m{integerNegate \mapsto SMT.integerAdd} fails this test: it replaces a unary function with a binary one, and an evaluator that calls \m{SMT.integerAdd} in place of \m{integerNegate} gets stuck.
The embedding \m{integerAdd \mapsto SMT.integerAdd} has the right arity but still the wrong type: the original has type \m{Integer \rightarrow Integer \rightarrow Integer}, the interpretation \m{SMT.Integer \rightarrow SMT.Integer \rightarrow SMT.Integer}.
The type embedding closes the gap.
Its axiom casts the solver's function to the type of the original:
\[
\m{integerAdd \mapsto \cast{SMT.integerAdd}{sym\ (C_{Integer} \rightarrow C_{Integer} \rightarrow C_{Integer})}}
\]
The coercion \m{C_{Integer} \rightarrow C_{Integer} \rightarrow C_{Integer}} equates the type of the original with the type of the solver's function, and \m{sym} turns it around.
%

\mypara{Unique Representation}
The axiom \m{C_{Integer}} says that \m{Integer} and \m{SMT.Integer} are the same type.
In reality, they are not: GHC defines \m{Integer} as a data type whose constructor \m{IS} wraps a machine integer.
After the embedding above, \m{\app{\app{integerAdd}{(\app{IS}{1})}}{n}} hands the constructor \m{\app{IS}{1}} to \m{SMT.integerAdd}, which expects a number, and the type checker cannot object.
We therefore give an embedded type a unique representation, its interpretation, by \emph{invalidating} the original: every constructor of an embedded type becomes \m{\ub}, and so does the axiom of an embedded \lstinline{newtype}.
A run that reaches \m{\ub} has found an operation on the original type that the
embedding does not cover.

\mypara{Coherence}
Type embeddings should not overlap.
Consider the two embeddings \m{[\alpha] \mapsto \app{SMT.Seq}{\alpha}} and \m{[Char] \mapsto SMT.String}.
Each is reasonable alone: the first models lists with the theory of sequences, the second lists of characters with the theory of strings.
Together they give \m{[Char]} two interpretations, and their axioms compose into
a coercion between \m{\app{SMT.Seq}{Char}} and \m{SMT.String}, which the solver
needs to distinguish.
We thus require that no two original types unify.

\mypara{Well-Formed Embeddings}
An embedding \m{\embed} is a pair of a set of type embeddings \m{\mathcal{A}} and a map of term embeddings \m{\rho}:
\[
\m{\embed \bnf (\mathcal{A}, \rho)}
\hspace{30pt}
\m{\mathcal{A} \bnf \{ C : \quant{\vect{\alpha}}{\app{T}{\vect{\tyInst}} \sim \tyInst'} \}}
\hspace{30pt}
\m{\rho \bnf \{ x \mapsto t \}}
\]
Each type embedding is an axiom whose original \m{\app{T}{\vect{\tyInst}}} is a type constructor applied to arguments.
We say that \m{\mathcal{A}} \emph{embeds} \m{T}.
Each term embedding maps a top-level variable to a thunk: the interpretation,
paired with an environment that suffices to evaluate it.
%
%
\Cref{fig:wf} defines when an embedding is well formed, written \m{\vdash \embed}.
Rule \m{[Coh]} is the coherence check: every axiom relates two types of the same kind, and no two originals unify.
Here \m{unify(\tyInst_1, \tyInst_2) = \bot} says that no substitution of types for type variables makes \m{\tyInst_1} and \m{\tyInst_2} equal.
Rule \m{[Emb]} adds the term embeddings: each interpretation has the type of its original.
We write \m{x : \tyInst} when variable \m{x} has type \m{\tyInst} in the program, and \m{\mathcal{A} \vdash t : \tyInst} when thunk \m{t} has type \m{\tyInst} in \FC extended with the axioms \m{\mathcal{A}}, under the restriction on embedded type constructors above (\Cref{def:typing}).

\begin{figure}[t]
\centering
\begin{minipage}[b]{0.58\linewidth}
\begin{mathpar}
\inferrule*[right=\m{Coh}]
  { \m{\forall (C : \quant{\vect{\alpha}}{\app{T}{\vect{\tyInst}} \sim \tyInst'}) \in \mathcal{A}.\ \
      \app{T}{\vect{\tyInst}} : \kappa \ \land\ \tyInst' : \kappa}
    \\\\
    {\begin{array}{@{}l@{}}
      \m{\forall (C_1 : \quant{\vect{\alpha_1}}{\tyInst_1 \sim \tyInst_1'}),\ (C_2 : \quant{\vect{\alpha_2}}{\tyInst_2 \sim \tyInst_2'}) \in \mathcal{A}.} \\
      \m{\quad C_1 \neq C_2 \implies unify(\tyInst_1, \tyInst_2) = \bot}
    \end{array}}}
  {\m{\vdash \mathcal{A}}}
\end{mathpar}
\end{minipage}
\hfill
\begin{minipage}[b]{0.4\linewidth}
\begin{mathpar}
\inferrule*[right=\m{Emb}]
  { \m{\vdash \mathcal{A}}
    \\ \m{\forall (x \mapsto t) \in \rho.\ \ x : \tyInst \implies \mathcal{A} \vdash t : \tyInst}}
  {\m{\vdash (\mathcal{A}, \rho)}}
\end{mathpar}
\end{minipage}
\caption{Well-formed embeddings. \label{fig:wf}}
\end{figure}

\begin{figure}[t]
\centering
\[
\begin{array}{lcll}
\m{invalidate_{\mathcal{A}}(D)} & \m{=} & \m{\ub}
  & \text{if } \m{D} \text{ is stale} \\
\m{invalidate_{\mathcal{A}}(\coInst)} & \m{=} & \m{\ub}
  & \text{if } \m{\coInst} \text{ is stale} \\
\m{invalidate_{\mathcal{A}}(\cast{e}{\coInst})} & \m{=} & \m{\ub}
  & \text{if } \m{\coInst} \text{ is stale} \\
\m{invalidate_{\mathcal{A}}(\cast{e}{\coInst})} & \m{=} & \m{\cast{invalidate_{\mathcal{A}}(e)}{\coInst}}
  & \text{otherwise} \\
\m{invalidate_{\mathcal{A}}(\abs{x}{e})} & \m{=} & \m{\abs{x}{invalidate_{\mathcal{A}}(e)}} \\
\m{invalidate_{\mathcal{A}}(\app{e_f}{e_a})} & \m{=} & \m{\app{invalidate_{\mathcal{A}}(e_f)}{invalidate_{\mathcal{A}}(e_a)}} \\
\m{invalidate_{\mathcal{A}}(\case{e}{\alts{a}})} & \m{=} & \m{\case{invalidate_{\mathcal{A}}(e)}{\vect{invalidate_{\mathcal{A}}(a)}}} \\
\m{invalidate_{\mathcal{A}}(\alt{D}{\vect{x}}{e})} & \m{=} & \m{\alt{D}{\vect{x}}{invalidate_{\mathcal{A}}(e)}} \\
\m{invalidate_{\mathcal{A}}(\app{raise}{e})} & \m{=} & \m{\app{raise}{invalidate_{\mathcal{A}}(e)}} \\
\m{invalidate_{\mathcal{A}}(e)} & \m{=} & \m{e}
  & \text{otherwise}
\end{array}
\]
\[
\begin{array}{lcl}
\m{invalidate_{\mathcal{A}}(\thunk{\subst}{e})} & \m{=} &
  \m{\thunk{invalidate_{\mathcal{A}}(\subst)}{invalidate_{\mathcal{A}}(e)}} \\
\m{invalidate_{\mathcal{A}}(\subst)} & \m{=} &
  \m{\{ x \mapsto invalidate_{\mathcal{A}}(t) \mid (x \mapsto t) \in \subst \}} \\[6pt]
\m{remap(\embed, \subst)} & \m{=} &
  \m{\{ x \mapsto embed(\embed, t) \mid (x \mapsto t) \in \subst,\ x \notin dom(\rho) \} \cup invalidate_{\mathcal{A}}(\rho)} \\
\m{embed(\embed, \thunk{\subst}{e})} & \m{=} &
  \m{\thunk{remap(\embed, \subst)}{invalidate_{\mathcal{A}}(e)}}
\end{array}
\]
\caption{Applying a well-formed embedding \m{\embed = (\mathcal{A}, \rho)} to a thunk. \label{fig:embed}}
\end{figure}

\mypara{Translation}
\Cref{fig:embed} defines the translation \m{embed(\embed, t)}, which applies a
well-formed embedding \m{\embed} to a \core thunk \m{t} and returns another
\core thunk.
Function \m{invalidate_{\mathcal{A}}} removes the original representation from an expression, and from a thunk or an environment by visiting every expression in it.
A data constructor is \emph{stale} if \m{\mathcal{A}} embeds its type constructor.
A coercion is stale if it mentions the axiom of a \lstinline{newtype} that
\m{\mathcal{A}} embeds, or if it breaks the restriction on embedded type
constructors above; We give the full definition of \Cref{def:stale} in
\appcref{app:embed}.
A stale constructor, a stale coercion and any cast that uses one all become
\m{\ub}.
Everywhere else, the function descends into the expression and leaves it as it is; in particular, it keeps the patterns of a case split.

\noindent
Function \m{remap} replaces definitions in an environment.
The interpretations in \m{\rho} override the bindings of the same name, and every other binding is translated with \m{embed}.
We assume that the embedded variables differ from the variables that abstractions and patterns bind.

\mypara{Guarantees}
Let \m{\embed = (\mathcal{A}, \rho)} be well formed, and let \m{t} be a \core thunk.
The translation keeps the program well typed.

\begin{theorem}[Type preservation]
\label{thm:embed-typing}
If \m{t} has type \m{\tyInst}, then \m{\mathcal{A} \vdash embed(\embed, t) : \tyInst}.
\end{theorem}

\noindent
No coercion after the translation equates two different data types, where a
\emph{data type} is a type whose head is the function constructor, or the type
constructor of a data type that \m{\mathcal{A}} does not embed.

\begin{theorem}[Consistency]
\label{thm:embed-consistent}
Let \m{\coInst} be a closed coercion that is not stale.
If \m{\coInst : \tyInst_1 \sim \tyInst_2}, and \m{\tyInst_1} and \m{\tyInst_2} are data types, then \m{\tyInst_1} and \m{\tyInst_2} have the same head.
\end{theorem}


%
\noindent
\appCref{app:embed} sketches the proofs.
Finally, the translated thunk is a \core program, so \Cref{thm:soundness,thm:completeness} apply to it.

%


\section{Implementation}
\label{sec:impl}

\tool is a full-path symbolic evaluation for GHC Core, implemented as a GHC plugin.
\tool is written in $\sim$8000 lines of Haskell code, and uses the Grisette
library for both its ordered guards merging algorithm and its
primitive symbolic data types.

\subsection{Evaluator}
\label{sec:impl:eval}
\mypara{Types, Casts and Coercions}
\FC permits syntactic erasure of types and casts at runtime, and represents
coercions as zero-sized unit values.
\tool follows suit and directly strips types, type abstractions, type
applications and casts from expressions, and represents coercions as unit values
during symbolic evaluation.

\mypara{Primitives}
\tool exposes SMT primitives via a set of built-in types and operations.
\begin{lstlisting}
data SMT.Bool = ...
data SMT.BitVec (n :: Nat) = ...
data SMT.Array k v = ...

SMT.and :: SMT.Bool -> SMT.Bool -> SMT.Bool
SMT.bvadd :: SMT.BitVec n -> SMT.BitVec n -> SMT.BitVec n
SMT.select :: SMT.Array k v -> k -> v
\end{lstlisting}
The functions have a built-in interpretation in \tool which binds them to their
respective SMT operations during symbolic evaluation.
We also provide a direct Haskell implementation, which allows the user to
execute the primitives when receiving a satisfying instance from the solver.
Exposing primitives in this way rather than deploying them through Haskell
primitive types is a deliberate design choice.
It is, for example, more convenient to have \lstinline{SMT.BitVec} available
directly compared to \lstinline{Int8#}, \lstinline{Int16#}, etc.
In fact, even primitive types might require a different embedding based on the
verification task, which is possible in our design.

\mypara{Primitive Optimisation}
To keep symbolic evaluation fast, \tool employs lightweight, but effective
rewrites of primitive types, which it inherits from Grisette.
This technique is not new, but effectively reduces constraint size.
For example, we fold constant by rewriting \lstinline{SMT.bvadd 6 5} \m{\mapsto}
\lstinline{0xb}.
As ordered guards style merging often introduces large compound conditionals, we
find that boolean rewrites such as \m{x \land (\neg x\lor y) \mapsto x \land y}
are vital to keep branch conditions small.

\mypara{Literals}
\tool permits Haskell primitive types such as \lstinline{Int64#} to be embedded as any arbitrary type.
This freedom makes the translation of literals in Haskell ambiguous.
Consider an example where we define \lstinline{four = 4#Int64} as a constant of type \lstinline{Int64#}.
The translation is dependent on the type embedding of \lstinline{Int64#}.
To resolve this ambiguity, \tool exposes a translation function for each
literal, in this case \lstinline{fromLitInt64# :: SMT.BitVec 64 -> Int64#}.
For this, we pick a lossless representation for each literal:
\lstinline{SMT.BitVec 64}.
An embedding of \lstinline{fromLitInt64#} describes how to convert into an \lstinline{Int64#} using the appropriate type embedding.
When \tool evaluates a literal, we call its respective translation function with the lossless representation as argument.
Of course, we expect almost all use cases to want to embed the sized integer and word primitives of Haskell to the bit vector of their respective size, e.g. \lstinline{Int64#} \m{\mapsto} \lstinline{SMT.BitVec 64}.
For this purpose, \tool exposes a set of standard type and term embeddings.

\subsection{Embeddings}
\label{sec:impl:embed}
Embeddings are local to each invocation as every verification task might require different, possibly incompatible embeddings.
For this purpose, \tool exposes data type \lstinline{Embedding}.
An embedding is just a normal top-level Haskell value, and an annotation refers to it by name.
\tool represents a type embedding as a tuple containing the original type and the interpretation type.
We use the same representation for term embeddings, but over variables instead of types.
\begin{lstlisting}
data Embedding = Embedding
  { typeEmbeddings :: [(Type, Type)]
  , termEmbeddings :: [(Var, Var)] }
\end{lstlisting}
Embeddings in \tool are transformations applied to GHC Core, not \calc: this
lets us write embeddings in Haskell.
We translate from GHC Core to \calc after resolving embeddings.

\mypara{Embeddable}
Type embeddings require both types to have the same kind, which rules out primitive types: \lstinline{Int64#} has kind \lstinline{TYPE Int64Rep}, while \lstinline{SMT.BitVec 64} has kind \lstinline{TYPE LiftedRep}.
\tool relaxes this constraint and lets the runtime representation of embeddings
differ.
As Haskell's \lstinline{Coercible} fixes its two types to one kind, \tool
provides its own typeclass \lstinline{Embeddable} for coercions between types
whose kinds differ in this way; we describe this in more detail in
\appcref{app:embeddable}.

\mypara{Term Embeddings}
A term embedding can consume an \lstinline{Embeddable} instance through its type signature.
Below we define an embedding for \lstinline{plusInt64#}:
\begin{lstlisting}
plusEmbed :: Embeddable Int64# (SMT.BitVec 64)
          => Int64# -> Int64# -> Int64#
plusEmbed = coerce $ SMT.bvadd @64
\end{lstlisting}
\tool locally provides an \lstinline{Embeddable} instance to a term embedding if it can be resolved from its type embeddings.
The term embedding is accepted if the final type signature matches the original after the resolution of \lstinline{Embeddable} constraints.

\mypara{Lightweight}
The embedding functionality in \tool is extremely lightweight.
It is implemented in $\sim$500 lines of Haskell code.
This is because almost all functionality can reuse the existing GHC codebase for their implementation.
Users do not need to learn a new language to write embeddings: instead their definition can even sit right next to the code they verify.

\section{Case Study}
\label{sec:case-study}
Next, we discuss the embedding of \lstinline{IORef}, which provides safe, mutable
pointers in Haskell.
Though \lstinline{IORef} exposes only three primitive operations, other IO
operations can be embedded in the same way.

\mypara{IO Monad}
To avoid impurity in the presence of lazyness, Haskell uses a simple unit
datatype called \lstinline{RealWorld#}, whose defining feature is that there can
only a single instance per execution.
The idea is straightforward: an operation can only perform IO once it gets the
\lstinline{RealWorld#} token.
Once the operation is finished, it passes the token to the next operation, which
can then, in turn, perform IO.
Funtion \lstinline{getChar} which gets a char from user input is defined as
\lstinline{getChar :: RealWorld# -> (RealWorld#, Char)}.
With this, we can define function \lstinline{get3Char} that fetches three chars
from user input, in order.
\begin{lstlisting}
get3Char :: RealWorld# -> (RealWorld#, [Char])
get3Char w0 = let (w1, c0) = getChar w0 ; let (w2, c1) = getChar w1
              let (w3, c2) = getChar w2   in (w3, [c0, c1, c2])
\end{lstlisting}
Passing around the \lstinline{RealWorld#} token is tedious, so Haskell abstracts
token passing in the \lstinline{IO} monad, defined as
\lstinline{newtype IO a = IO (RealWorld# -> (RealWorld#, a))}.
The monadic bind operation \lstinline{(>>=)} performs the token passing as we did manually in \lstinline{get3Char}.

\mypara{Embed RealWorld\#}
In \tool, our goal is to model impure computation during symbolic evaluation.
The core idea is simple: instead of its original unit token, our embedding of
\lstinline{RealWorld#} contain the outside world, defined as data type
\lstinline{ModelWorld}.
The contents of \lstinline{ModelWorld} follow from the IO operations we are
trying to describe.
For now, we will restrict ourselves to modeling the \lstinline{IORef} data type
and its operations, which is responsible for safe, mutable pointers in Haskell.

\mypara{IORef}
The \lstinline{IORef a} data type is internally a pointer to the heap storing a value of type \lstinline{a}.
It supports three primitive operations which allows us to create,
read from and write to an \lstinline{IORef}, respectively:
\begin{lstlisting}
newIORef :: a -> IO (IORef a)
readIORef :: IORef a -> IO a
writeIORef :: IORef a -> a -> IO ()
\end{lstlisting}
For example, the following program initializes an \lstinline{IORef} with Integer
\lstinline{5}, reads it, increments its value, and finally re-read it.
\begin{lstlisting}
main = do
  ref <- newIORef (5 :: Int); val0 <- readIORef ref -- val0 == 5
  writeIORef ref (val0 + 1);  val1 <- readIORef ref -- val1 == 6
\end{lstlisting}

\mypara{Heap Model}
To model \lstinline{IORef}, we need to support allocations inside of \lstinline{ModelWorld}.
We can model this with an \lstinline{IntMap Any} that tracks allocations (\ie a
model heap).
To support heterogenous allocations, we store the values as an \lstinline{Any}
type: an opaque type that may be coerced to from any other type.
We coerce values of type \lstinline{Any} back to their original type during
reads of an \lstinline{IORef}.
Every allocation gets a new unique allocation identifier (\ie a pointer) of type
\lstinline{Int}.
We generate these identifiers by reading and incrementing the current size of
the heap, in the style of a bump allocator.
The identifiers as well as the heap itself remain concrete, which keeps the
lookups fast.
We define the heap as \lstinline{newtype ModelWorld = ModelWorld (IntMap Any)}.
We embed the type \lstinline{IORef a} as an \lstinline{Int}: this is the pointer
into our heap.
With these type embeddings, we can implement the term embeddings for functions
\lstinline{newIORef}, \lstinline{readIORef} and \lstinline{writeIORef}.

\mypara{Allocation}
We start with \lstinline{newIORef}.
We fetch a unique identifier, which we obtain by taking the length of the
current heap.
After this, we insert the initial element for this entry into the heap and
return the updated world, together with the identifier that references the
entry.
Note that we erase the type of the initial value by coercing it using \lstinline{toAny}.
We define a function of type \lstinline{IntMap Any -> (IntMap Any, Int)} which  takes the current heap and produces a new heap and identifier.
As both \lstinline{IO} and \lstinline{ModelWorld} are newtypes, we can \lstinline{coerce} this expression directly to the required type \lstinline{IO (IORef a)}.
\begin{lstlisting}
newIORefE :: Embeddable RealWorld# ModelWorld
          => Embeddable (IORef a) Int => a -> IO (IORef a)
newIORefE init = coerce \heap0 ->
  let ref = length heap0; heap1 = insert ref (toAny init) heap0
  in (heap1, ref)
\end{lstlisting}

\mypara{Read}
Next, we embed \lstinline{readIORef}.
To support reads, we have to reify the type of a value in our heap.
That is, our heap stores heterogeneous values of the opaque type \lstinline{Any}, which is achieved by casting the original values to this type.
This coercion is only safe because \lstinline{Any} supports no operations.
Coercing back to the original type however requires the usage of \lstinline{unsafeCoerce}, which, as the name suggests, conjures an unsafe coercion witness to perform a cast.
Fortunately, the type parameter on \lstinline{IORef a} provides a safety mechanism: every \lstinline{IORef a} allocates a value of type \lstinline{a}.
As such, this also holds true for the unique identifier \lstinline{Int} as which an \lstinline{IORef a} is embedded.
With this, we can safely apply \lstinline{unsafeCoerce} in our embedding of \lstinline{readIORef}.
The rest of the embedding is a basic lookup from the heap.
The heap itself remains unchanged.
\begin{lstlisting}
readIORefE :: Embeddable RealWorld# ModelWorld
           => Embeddable (IORef a) Int => IORef a -> IO a
readIORefE = coerce \ref heap ->
  let elem0 = lookup ref heap :: Any; elem1 = unsafeCoerce elem0 :: a
  in (heap, elem1)
\end{lstlisting}

\mypara{Write}
Finally, the embedding for \lstinline{writeIORef} is rather straightforward: it is only an update of the heap as we already showed in \lstinline{newIORefE}.
\begin{lstlisting}
writeIORefE :: Embeddable RealWorld# ModelWorld
            => Embeddable (IORef a) => Int -> IORef a -> a -> IO ()
writeIORefE = coerce \ref value heap0 ->
  let heap1 = insert ref (toAny value) heap0 in (heap1, ())
\end{lstlisting}

\mypara{Embedding}
All the embeddings we discussed need to be provided to an invocation of \tool.
The following \lstinline{Embedding} data structure contains the embeddings required for \lstinline{IORef} support, which also doubles as a summary of the case study.
\begin{lstlisting}
embeddingIORef :: Embedding
embeddingIORef = Embedding
  { typeEmbeddings = [ RealWorld# %$\mapsto$% ModelWorld, IORef a %$\mapsto$% Int]
  , termEmbeddings = [ newIORef %$\mapsto$% newIORefE, readIORef %$\mapsto$% readIORefE
                     , writeIORef %$\mapsto$% writeIORefE ] }
\end{lstlisting}
Interestingly, \lstinline{ModelWorld} only contains the non-symbolic Haskell
data type \lstinline{IntMap}.
This shows that embeddings are useful, not just for SMT Theories, but for any
data type.
The same heap also supports raw, byte-addressable pointers
(\lstinline{Ptr} and \lstinline{ForeignPtr}) as used by libraries such as
\texttt{text} and \texttt{bytestring}; we describe this embedding in
\appcref{app:ptr}.

\section{Evaluation}
\label{sec:evaluation}
%

\subsection{G2}
We compare \tool against G2 using the benchmarks from G2Q \cite{G2Q}, the quasi
quoter interface of G2.
The G2 benchmark consists of four programs: an imperative arithmetic language, a
program synthesizer for the lambda calculus, an n-queens solver and a solver to
find matching strings of a regular expression.
We additionally reproduce the results of Grisette \cite{Grisette} on the same
benchmarks, as a baseline comparison.
Grisette runs as a lightweight library in compiled Haskell; porting the
benchmarks requires rewriting the benchmarks within the Grisette library,
directly using its symbolic primitives like \lstinline{Union} for representing
case-splits. This port was already performed by the Grisette authors for their own
benchmarks, so we reuse it.
\tool directly interprets the program as is, by compiling it to GHC core and
interpreting the resulting expression symbolically.
As \tool uses Grisette internally, Grisette's performance quantifies the
overhead of symbolically interpreting compiled GHC core.
\cref{tab:g2q} shows the performance of G2, \tool and Grisette.
We ran the benchmarks with a timeout cap of 90 seconds.

\mypara{Search}
G2 performs a breadth-first search over symbolic paths to find
counterexamples.
This allows them find counterexamples in programs with un-boundedly many paths.
\tool does not have this built-in search procedure, but we can replicate it via
iterative deepening. Grisette's port of the benchmarks uses the same approach.

\mypara{Performance}
\tool is faster than G2 on most benchmarks, with a geometric mean speedup of
6.25x.
G2 outperforms \tool on benchmarks with very low runtime, such as
\lstinline{solveDeBruijn}.
We suspect that most of the time spent by \tool goes to the
preparation of embeddings.
We use \tool and its \clash embeddings (\appcref{app:clash}) to verify the leakage description of
AIMCore~\cite{10.1145/3808340}, a 5-stage in-order RISC-V processor implemented
in approximately 800~lines of \clash.
While in the worst case theoretically, choosing between all-paths and single-path symbolic
execution means choosing between exponential memory and exponential time;
in our experience, however, state merging works well on Haskell code in practice: it keeps
the symbolic state small, and so saves both time and memory.
For example the leakage proof of AIMCore takes 11.5\,s, and its memory use peaks at only 441\,MB,
but with state merging disabled, the proof times out after 10 minutes.
This demonstrates that for realistic code, state merging prevents catastrophic
runtime explosion while keeping memory overhead modest and practical.

\mypara{Benchmark Difference}
We note that the benchmark results of Grisette on \lstinline{solveDeBruijn}
differ from what is reported in the original paper.
The artifact of Grisette restricted the symbolic value generation only to closed DeBruijn terms, which greatly reduces statespace.
For a fair comparison with G2, we adjusted the benchmark to be more faithful to
the original formulation.
%

\begin{figure}[t]
\centering\footnotesize
\setlength{\tabcolsep}{3pt}
\begin{minipage}[t]{0.53\linewidth}
\centering
\begin{tabular}{@{}lrrr@{}}
\toprule
Task & G2 & \tool & Grisette \\
\midrule
\texttt{badEnvSearch}                  & 59.32 & 0.09 & 0.02 \\
\texttt{Search Mul=Add}                & 0.56 & 0.21 & 0.10 \\
\texttt{DeBruijn id/const} (2)       & 0.04--1.06 & 0.07--0.08 & 0.02--0.04 \\
\texttt{DeBruijn OR}                   & 86.22 & 0.27 & 0.03 \\
\texttt{DeBruijn NOT}                  & Timeout & Timeout & Timeout \\
\texttt{DeBruijn AND}                  & Timeout & 20.52 & 12.32 \\
\texttt{Queens 4}--\texttt{8} (5)      & 0.36--2.30 & 0.08--0.16 & 0.03--0.04 \\
\texttt{stringSearch} (4)              & 0.02--4.23 & 0.09--0.12 & 0.02--0.12 \\
\bottomrule
\end{tabular}
\captionof{figure}{G2Q Benchmarks in seconds.}
\label{tab:g2q}
\end{minipage}
\hfill
\begin{minipage}[t]{0.44\linewidth}
\centering
\begin{tabular}{@{}lrrrr@{}}
\toprule
Domain & Props & \cmark & \xmark & ms \\
\midrule
\texttt{Bool}                      & 2  & 1  & 1  & 8--25 \\
\texttt{Integer}                   & 2  & 2  & 0  & 5--6 \\
\texttt{Int}, \texttt{Int8}--\texttt{Int64} & 13 & 8 & 5 & 5--10 \\
\texttt{Word}, \texttt{Word8}, \texttt{Word64} & 8 & 5 & 3 & 5--9 \\
\texttt{Data.Map.Strict} & 10 & 7 & 3 & 5--12 \\
\texttt{Data.Set} & 5 & 4 & 1 & 4--5 \\
\texttt{Data.IP} & 7  & 7  & 0  & 5--17 \\
\texttt{IORef}                     & 1  & 1  & 0  & 7 \\
\texttt{Ptr} / \texttt{ForeignPtr} & 6 & 6 & 0 & 6--7 \\
\midrule
Total                                 & 54 & 41 & 13 & 389 \\
\bottomrule
\end{tabular}
\captionof{table}{Properties by domain: \cmark\ valid, \xmark\ invalid.}
\label{tab:properties}
\end{minipage}
\end{figure}

\subsection{Verified Properties}

We evaluate \sys on a suite of 54 properties spanning primitive types, standard
library data structures, real-world third-party libraries, and low-level
pointer and memory operations.
Each property is verified at compile time by the GHC plugin, which invokes Z3;
a property that gets an unexpected result stops compilation.
Of these, 13 are deliberately invalid (\cref{tab:properties}),
spread across all type domains to exercize \sys's counterexample generation
end-to-end.
All experiments are run on an Apple M4 Pro with 24\,GB of RAM\@.
\Cref{tab:properties} summarises the properties per domain together with their
SMT solve times as reported by Z3. The full tables are in \appcref{app:properties}.

\mypara{Primitive Types}
We verify arithmetic properties of \lstinline{Bool}, \lstinline{Integer}, the
signed integers \lstinline{Int}, \lstinline{Int8}, \lstinline{Int16},
\lstinline{Int32}, \lstinline{Int64}, and the unsigned types \lstinline{Word},
\lstinline{Word8}, \lstinline{Word64}.
They include commutativity and identity of addition, self-subtraction to zero,
commutativity of multiplication, bitwise-AND commutativity for
\lstinline{Word}, De Morgan's law for \lstinline{Bool}, and $x < x + 1$ for
unbounded \lstinline{Integer}.
Each type also has an invalid property such as $x < x$.

\mypara{Data.Map.Strict and Data.Set}
We verify 15 properties of \lstinline{Data.Map.Strict} and \lstinline{Data.Set}
from the \texttt{containers} package, which ships with GHC.
They cover membership and lookup after construction with \lstinline{empty},
\lstinline{singleton}, \lstinline{insert} and \lstinline{delete}:
deleting the only key of a singleton yields a null map; and a key distinct from
the singleton key is not a member.
The four invalid properties cover membership after self-deletion and lookup
returning \lstinline{Nothing} after an overwrite.

\mypara{iproute}
We verify 8 properties of address arithmetic from the external library
\texttt{iproute}.
The byte-swap helper \lstinline{fixByteOrder} is an involution.
\lstinline{makeAddrRange} always contains its base address.
Subnet containment (\lstinline{>:>}) is reflexive for IPv4 and IPv6.
For IPv4 it is also transitive, and it agrees with the matching IPv4-mapped
IPv6 range.
Transitivity is the hardest property: it has six arguments and a precondition
that bounds mask lengths to $[0, 32]$.
Verification also exposed a bug in \lstinline{maskIPv6}.
For the mask length \lstinline{minBound :: Int}, \lstinline{negate} overflows
back to \lstinline{minBound}, and \lstinline{shiftR} then receives a negative
shift count and fails at runtime.
We work around it with \lstinline{Word8}.

\mypara{IORef and Ptr / ForeignPtr}
We verify 7 properties of mutable memory.
For \lstinline{IORef}, a value written to a reference and read back equals the
original, which validates the IO embedding of \cref{sec:case-study}.
For \lstinline{Ptr} and \lstinline{ForeignPtr}, 6 properties cover raw, unsafe
memory access against the byte-addressable symbolic heap (\appcref{app:ptr}):
a fresh allocation reads as zero;
a write at one symbolic offset does not alias a read at a distinct offset;
a second write overwrites the first;
a write through nested pointer arithmetic is seen at the equivalent flat offset;
a write to one allocation does not alias a read from another; and finally
a write is still seen after \lstinline{castPtr} to another element type and
back.

\section{Limitations}



\mypara{Primitive Containers}
SMT exposed container types like \lstinline{SMT.Array} can only reason about
other SMT primitives that it contains.
For example, we can reason about \lstinline{SMT.Integer} as both an index or an
element of an \lstinline{SMT.Array}.
SMT containers do not however permit arbitrary types.
This is not a problem if the contained type is known: we can write a custom encoding of our data type expressed in SMT primitives.
A more problematic case is when we wish to have an embedding of a Haskell container,
e.g. \lstinline{Map k v} $\mapsto$ \lstinline{SMT.Array k v}.
As the function \lstinline{Map.insert} would be polymorphic over its key and value, there is no general translation from the key and value type into SMT primitives.
We also cannot constrain the inputs to include only types that are convertible
to SMT, as this would violate the subsumption requirement of the term embedding.
As of writing, we do not have a satisfying way to support SMT container embeddings over unrestricted, polymorphic elements.

\section{Related Work}
Prior work develops symbolic evaluation with guarded unions and structural
merging~\cite{Rosette}, type systems for safely combining concrete and symbolic
operations~\cite{symbolic-types}, and a mechanized correctness framework for
evaluation with merging~\cite{symbolic-merging}. These calculi use eager
evaluation. Our calculus instead extends a lazy semantics based on System FC,
with guarded alternatives over both values and thunks: merging matching
constructors preserves suspended fields, which are forced only when demanded.
Although \citet{symbolic-merging} also express assumptions and assertions
through conditionals and abort/error constructs, we retain their outcomes within
guarded values and thunks, without a separate assumption/assertion store. This
integrates symbolic branching, merging, and failure propagation with ordinary
lazy evaluation.

Our implementation builds on Grisette~\cite{Grisette}, a Haskell library for
symbolic execution, for its ordered guard representation and merging of symbolic
values. Whereas using Grisette requires expressing programs with its symbolic
library operations, \tool directly interprets native Haskell programs in GHC
Core.

Hs-to-coq~\cite{10.1145/3236784} translates Haskell into the Rocq proof
assistant~\cite{rocq_prover} for foundational proofs. Liquid
Haskell~\cite{liquidhaskell} uses refinement types, while
Stainless~\cite{stainless} verifies Scala function contracts using SMT.
\tool instead checks executable Haskell properties through bounded symbolic
evaluation with configurable representations.

Soft contract verification~\cite{scv-oopsla12,scv-icfp14,scv-pldi15,scv-popl18}
and Crux-MIR~\cite{crux_galois} check properties stated as ordinary
host-language functions: executable Racket predicates and Rust functions, much
like our properties. Crux-MIR uses Crucible's all-paths symbolic execution with
state merging.
Unlike \tool, both fix the symbolic semantics of primitives, data and state in
the analysis; Crux-MIR supports standard-library code through developer-curated
overrides, and adding primitive types requires extending the engine.
In \tool it is configurable via embeddings written in Haskell, so property
authors can choose representations per property without modifying the engine;
\eg our support for \lstinline{IORef} and \lstinline{Ptr} is ordinary user code
rather than an extension of the engine.
G2~\cite{lazycounter} performs symbolic execution in Haskell. G2 offers
counterfactual symbolic execution which allows detecting errors that do not
correspond to concrete property violations, but rather, stem from imprecision of
the type system. Unlike \tool, G2 performs King-style symbolic
execution~\cite{king} by symbolically evaluating individual paths guarded by a
path-condition. By contrast, we perform all-paths symbolic execution which
allows us to perform state merging, which can benefit
speed~\cite{merging,veritesting}. Unlike \tool, G2
does not directly represent arbitrary System FC conditions symbolically, and
cannot model arbitrary functions as other Haskell functions and SMT primitives
via embeddings.

Symbolic memory models offer different tradeoffs between precision and
scalability; we refer to Baldoni et al.~\cite{SurveySymExec-CSUR18} for a survey.
Our \lstinline{Ptr}/\lstinline{ForeignPtr} model (\appcref{app:ptr}) illustrates
how such modeling choices can be implemented as ordinary Haskell embeddings
without extending the symbolic evaluator.

\section*{AI Use}
\label{sec:ai-use}
We used Claude Code with Claude Opus 5.5 only for the Rocq mechanisation of \cref{sec:theory}.
The tool wrote the Rocq definitions and proof from the definitions in
\cref{sec:theory}, and the authors decided the design of the calculus.
The Rocq definitions and statements were checked manually against \cref{sec:theory}.
\section*{Data-Availability Statement}
The source code of \tool and the benchmarks of \Cref{sec:evaluation} are available at \url{https://github.com/PLSec-VU/symfc}.
The Rocq mechanisation of the theorems in \Cref{sec:theory} is available at \url{https://github.com/PLSec-VU/symfc-rocq}.

\bibliographystyle{ACM-Reference-Format}
\bibliography{references}

\appendix

\newcommand{\rocq}[1]{\textsf{#1}}

\section{Proof Sketches}
\label{app:proofs}
We mechanised \calc, \core and \Cref{thm:concrete-determinism,thm:soundness,thm:completeness}
in Rocq, and the development accompanies this paper.
It uses no axioms and admits no lemma.
This appendix gives the full instance relation and a sketch of each proof.
Each statement names its Rocq counterpart in \rocq{sans serif}.

\subsection{The Instance Relation}
\label{app:instances}

\begin{figure}[b]
\begin{mathpar}
\inferrule*[right=\m{I{-}Lit}]{ }{\m{\instance{l}{\sigma}{l}}}
\and
\inferrule*[right=\m{I{-}Sym}]{ }{\m{\instance{s}{\sigma}{\sigma(s)}}}
\and
\inferrule*[right=\m{I{-}Denote}]
  {\m{|\vect{p}| = arity(\prim)} \\ \m{\sigma(\app{\prim}{\vect{p}}) = l}}
  {\m{\instance{\app{\prim}{\vect{p}}}{\sigma}{l}}}
\\
\inferrule*[right=\m{I{-}Clos}]
  {\m{\instance{\subst_s}{\sigma}{\subst_c}} \\ \m{\instance{e_s}{\sigma}{e_c}}}
  {\m{\instance{\thunk{\subst_s}{\abs{x}{e_s}}}{\sigma}{\thunk{\subst_c}{\abs{x}{e_c}}}}}
\and
\inferrule*[right=\m{I{-}Con}]
  {\m{\instance{\vect{t_s}}{\sigma}{\vect{t_c}}}}
  {\m{\instance{\app{D}{\vect{t_s}}}{\sigma}{\app{D}{\vect{t_c}}}}}
\and
\inferrule*[right=\m{I{-}Partial}]
  {\m{\instance{\vect{t_s}}{\sigma}{\vect{t_c}}}}
  {\m{\instance{\app{\prim}{\vect{t_s}}}{\sigma}{\app{\prim}{\vect{t_c}}}}}
\and
\inferrule*[right=\m{I{-}Cast}]
  {\m{\instance{v_s}{\sigma}{v_c}}}
  {\m{\instance{\cast{v_s}{\coInst}}{\sigma}{\cast{v_c}{\coInst}}}}
\and
\inferrule*[right=\m{I{-}Atom}]
  {\m{v} \text{ is } \m{\tyInst} \text{ or } \m{\coInst}}
  {\m{\instance{v}{\sigma}{v}}}
\and
\inferrule*[right=\m{I{-}Empty}]{ }{\m{\instance{\cdot}{\sigma}{\cdot}}}
\\
\inferrule*[right=\m{I{-}IfT}]
  {\m{\sigma \models p} \\ \m{\instance{v_t}{\sigma}{v}}}
  {\m{\instance{\ife{p}{v_t}{v_f}}{\sigma}{v}}}
\and
\inferrule*[right=\m{I{-}IfF}]
  {\m{\sigma \models \neg p} \\ \m{\instance{v_f}{\sigma}{v}}}
  {\m{\instance{\ife{p}{v_t}{v_f}}{\sigma}{v}}}
\\
\inferrule*[right=\m{I{-}Susp}]
  {\m{\instance{\subst_s}{\sigma}{\subst_c}} \\ \m{\instance{e_s}{\sigma}{e_c}}}
  {\m{\instance{\thunk{\subst_s}{e_s}}{\sigma}{\thunk{\subst_c}{e_c}}}}
\and
\inferrule*[right=\m{I{-}TIfT}]
  {\m{\sigma \models p} \\ \m{\instance{t_t}{\sigma}{t}}}
  {\m{\instance{\ife{p}{t_t}{t_f}}{\sigma}{t}}}
\and
\inferrule*[right=\m{I{-}TIfF}]
  {\m{\sigma \models \neg p} \\ \m{\instance{t_f}{\sigma}{t}}}
  {\m{\instance{\ife{p}{t_t}{t_f}}{\sigma}{t}}}
\and
\inferrule*[right=\m{I{-}Extend}]
  {\m{\instance{t_s}{\sigma}{t_c}} \\ \m{\instance{\subst_s}{\sigma}{\subst_c}}}
  {\m{\instance{\subst_s\{x \mapsto t_s\}}{\sigma}{\subst_c\{x \mapsto t_c\}}}}
\end{mathpar}
\caption{The instance relation on values, thunks and environments.
On expressions, the relation is the congruence closure of
\m{\instance{s}{\sigma}{\sigma(s)}}.
On bottoms, it relates each bottom to itself, and \m{\app{raise}{e_s}} to
\m{\app{raise}{e_c}} when \m{\instance{e_s}{\sigma}{e_c}}.
\m{\instance{\vect{t_s}}{\sigma}{\vect{t_c}}} holds when the two sequences have equal
length and are related element by element.
\label{fig:instances}}
\end{figure}

\noindent
\Cref{fig:instances} gives the full relation (\rocq{contains\_val},
\rocq{contains\_thunk}, \rocq{contains\_env}, \rocq{contains\_expr} and
\rocq{contains\_prim}).
\m{I{-}Denote} is the only rule that relates a compound primitive value to a different
value, so the relation is semantic on SMT terms and syntactic elsewhere.
A purely syntactic relation would forbid \m{reduce{-}prim} to compute.
The branch rules inspect only the leaf that \m{\sigma} selects.

\subsection{Proofs}

\begin{lemma}[Merging keeps instances; \rocq{ite\_keeps}, \rocq{ite\_keeps\_false}, \rocq{merge\_keeps}]
\label{lem:ite-keeps}
If \m{\sigma \models p} and \m{\instance{v_t}{\sigma}{v}}, then
\m{\instance{ite(p, v_t, v_f)}{\sigma}{v}}.
If \m{\sigma \models \neg p} and \m{\instance{v_f}{\sigma}{v}}, then
\m{\instance{ite(p, v_t, v_f)}{\sigma}{v}}.
If \m{\instance{v_s}{\sigma}{v}}, then \m{\instance{merge(v_s)}{\sigma}{v}}.
\end{lemma}

\begin{proof}[Proof sketch]
By cases on the clauses of \Cref{fig:merge}, with induction for casts.
Two constructors or two partial applications merge into one whose thunks are branches,
and \m{I{-}TIfT} or \m{I{-}TIfF} picks the thunk that \m{\sigma} selects.
Two primitive values merge into the SMT term \m{\app{ite}{p\ p_t\ p_f}}, which
\m{I{-}Denote} relates to the literal that \m{\sigma} selects.
Two casts merge only under one coercion, so \m{I{-}Cast} applies to the merged bodies.
The last clause keeps the branch, and \m{I{-}IfT} or \m{I{-}IfF} applies.
\m{merge} is \m{ite} on a branch and the identity elsewhere.
\end{proof}

\begin{proof}[Proof sketch of \Cref{thm:concrete-determinism}]
In Rocq, the theorem is \rocq{concore\_eval\_deterministic}.
By one mutual induction on the first run, over \m{eval}, \m{apply}, \m{prim}, \m{force},
\m{select} and \m{fold{-}alts}.
The induction proves that each of these six relations is a function of its inputs on
\core arguments, at every bound.
A concrete run stays in \core, because \m{reduce{-}prim} computes a literal from literals
and \m{cast} keeps a \core value in \core.
So a concrete value is never a symbolic branch, and no rule that needs a branch applies.
\m{Prune} does not apply under \m{true}.
Any two of the remaining rules that match the same input have contradictory premises, so
both runs use the same rule, and the induction hypotheses make the results equal.
\end{proof}

\begin{proof}[Proof sketch of \Cref{thm:soundness}]
In Rocq, the theorem is \rocq{concore\_soundness}.
We prove a stronger statement (\rocq{lockstep\_sound}) by mutual induction on the
symbolic run.
For each of the six relations, a symbolic run at bound \m{n} on inputs that are related to
\core inputs gives a concrete run at the same bound \m{n}, and the two results are related.
The theorem takes \m{n = \infty}.
Most rules have a concrete counterpart with related premises.
Where the symbolic run splits on a branch in \m{apply}, \m{force}, \m{select} or
\m{fold{-}alts}, the concrete input is an instance of the leaf that \m{\sigma} selects.
Lifting over a branch spends no bound, so the induction hypothesis for that leaf gives a
concrete run at the same bound, and \Cref{lem:ite-keeps} relates the merged result.
The path condition of that leaf, \m{\pathc \land p} or \m{\pathc \land \neg p}, still
holds under \m{\sigma}.
So \m{\sigma} satisfies every path condition on the selected path, the solver accepts it,
and \m{Prune} never fires there.
For primitives, \m{reduce{-}prim} preserves instances, and applying it in each leaf of the
forced arguments keeps instances by \Cref{lem:ite-keeps} (\rocq{lift\_prim\_keeps}).
For \m{Type} and \m{Coercion}, substitution gives the same result in related
environments.
\end{proof}

\begin{lemma}[Runs without a bound are finite; \rocq{eval\_inf\_has\_budget}]
\label{lem:budget}
If \m{\bigstep{\pathc}{\subst}{e}{v}}, then there is an \m{h} such that
\m{\bigstepb{\pathc}{\subst}{n}{e}{v}} for every \m{n \ge h}.
\end{lemma}

\begin{proof}[Proof sketch]
By induction on the run.
A run is a finite derivation, and \m{h} is the largest number of nested \m{eval} steps in
it.
At every bound \m{n \ge h}, \m{Out{-}Of{-}Fuel} never applies, so the same derivation is
a run at bound \m{n}.
\end{proof}

\begin{proof}[Proof sketch of \Cref{thm:completeness}]
In Rocq, the theorem is \rocq{concore\_completeness\_budget}, and the form without
budget-totality is \rocq{concore\_completeness\_forall}.
First we prove the form without budget-totality.
\Cref{lem:budget} gives an \m{h} such that the concrete run of \m{e_c} gives \m{v_c} at
every bound \m{n \ge h}.
Then a mutual induction on the concrete run at bound \m{n} (\rocq{lockstep\_complete})
shows that every symbolic run at the same bound \m{n} on related inputs gives a related
result.
The induction mirrors the soundness proof.
At a branch, the concrete run follows the leaf that \m{\sigma} selects, and that leaf of
the symbolic run has the same bound, because lifting spends none.
A leaf that \m{\sigma} does not select may give \m{\budget} or any other value, and
\Cref{lem:ite-keeps} ignores it.
For the stated theorem, budget-totality gives a second bound \m{h'} at and above which a
symbolic run exists.
At every \m{n \ge \max(h, h')}, that run exists and its value is related to \m{v_c}.
\end{proof}

\mypara{Assumptions consistency}
A theorem about every solver that satisfies our assumptions says nothing if the assumptions
contradict each other.
We prove that a solver satisfying our assumptions exists (\rocq{model\_laws}).
It accepts every path condition, as \tool does during evaluation.
Its \m{reduce{-}prim} computes on literals and builds the SMT term
\m{\app{\prim}{\vect{p}}} otherwise, and its \m{cast} wraps the value in a cast.
We prove each assumption for it.
Inside it, we apply the three theorems to small programs, among them one that branches on
a symbolic Boolean and one that loops in a leaf the model does not select.

\mypara{Mechanisation notes}
The mechanisation follows \cref{sec:theory} with two presentational differences.
First, the bound is a fuel argument of all six relations, and only \m{eval} spends it.
Second, applying \m{reduce{-}prim} in each leaf of the forced arguments is a fixed
function, and only the step that computes the operation on primitive values is a
parameter of the solver.
This function gives \m{\ub} for a leaf that is not a primitive value, a bottom or a branch.

\section{Definitions for Embeddings}
\label{app:embed}
This appendix defines the notions that \Cref{sec:embed} uses informally: which constructors and coercions are stale, the typing judgement of \Cref{fig:wf}, and the data types of \Cref{thm:embed-consistent}.
Fix a \core program and a well-formed embedding \m{\embed = (\mathcal{A}, \rho)}.

\mypara{Signatures}
The program declares data types and \lstinline{newtype}s.
A data type \m{T} declares a set of constructors \m{cons(T)}, and a \lstinline{newtype} \m{N} declares one axiom \m{C_N : \quant{\vect{\alpha}}{\tyInst \sim \app{N}{\vect{\alpha}}}}.
These are the only axioms of the program.
The \emph{head} of a type is the type constructor at the root of its applications: \m{head(T) = T}, \m{head((\rightarrow)) = (\rightarrow)}, \m{head(\app{\tyInst_f}{\tyInst_a}) = head(\tyInst_f)} and \m{head(\cast{\tyInst}{\coInst}) = head(\tyInst)}; other types have no head.
The type constructors that \m{\mathcal{A}} embeds are
\m{emb(\mathcal{A}) = \{ T \mid (C : \quant{\vect{\alpha}}{\app{T}{\vect{\tyInst}} \sim \tyInst'}) \in \mathcal{A} \}}.

\mypara{Roles}
\core leaves out the roles of GHC's coercions, but we need them to say exactly which coercions an embedding rules out.
We follow Breitner et al.~\cite{SafeCoercions}.
A coercion is either nominal, written \m{\coInst : \tyInst_1 \ncoercion \tyInst_2}, or representational, written \m{\coInst : \tyInst_1 \rcoercion \tyInst_2}; a nominal coercion is also a representational one, and the axiom of a \lstinline{newtype} is representational.
Each parameter of a type constructor has a role.
A coercion \m{\app{T}{\vect{\coInst}}} between two applications of \m{T} is representational if each \m{\coInst_i} has at least the role of the \m{i}-th parameter of \m{T}, and the arguments of an application whose head is a type variable are nominal.
The coercions \m{left} and \m{right} apply to nominal coercions only.
The roles of parameters are inferred from the declarations: a parameter is nominal if a constructor field, or the representation of a \lstinline{newtype}, uses it in a nominal position, and representational otherwise.

\begin{definition}[Embedded signature]
\label{def:signature}
The signature of the embedded program differs from that of the program in three ways.
\begin{enumerate*}[label=(\arabic*)]
\item Every \lstinline{newtype} in \m{emb(\mathcal{A})} loses its axiom.
\item Every parameter of a type constructor in \m{emb(\mathcal{A})} is nominal, and the roles of all other parameters are inferred anew.
\item Every axiom of \m{\mathcal{A}} is added as a representational axiom.
\end{enumerate*}
\end{definition}

\noindent
Items (2) and (3) are how GHC treats a data family with one \lstinline{newtype} instance for each axiom of \m{\mathcal{A}}.
Inferring the roles anew matters, because the restriction on an embedded type constructor spreads to the types built from it.
If \m{\mathcal{A}} embeds \m{Maybe} and the program declares a data type \m{D} with a field of type \m{\app{Maybe}{\alpha}}, the parameter of \m{D} becomes nominal, and \m{\app{D}{Age}} is no longer related to \m{\app{D}{SMT.Integer}}.
The constructors of an embedded data type stay in the signature, for the patterns that mention them.

\begin{definition}[Stale]
\label{def:stale}
A data constructor \m{D} is \emph{stale} if \m{D \in cons(T)} for some \m{T \in emb(\mathcal{A})}.
A coercion is \emph{stale} if it is not well formed in the embedded signature.
\end{definition}

\noindent
A coercion of the program is thus stale in exactly two cases: it mentions the axiom of a \lstinline{newtype} in \m{emb(\mathcal{A})}, or it applies a type constructor to a representational coercion in a position that is nominal in the embedded signature.
The second case covers \m{\app{Maybe}{C_{MkAge}}} when \m{\mathcal{A}} embeds \m{Maybe}, and \m{\app{D}{C_{MkAge}}} for the type \m{D} above.
No use of \m{left} or \m{right} becomes stale: both apply to nominal coercions, which mention no axiom and so are well formed in either signature.
We assume that the casts inside types, which relate kinds, use no stale coercion.

\begin{definition}[Typing]
\label{def:typing}
We write \m{\mathcal{A}; \mathrm{\Theta} \vdash e : \tyInst} when expression \m{e} has type \m{\tyInst} in \FC with roles, in the embedded signature and under the typing context \m{\mathrm{\Theta}}.
A thunk has a type if its expression has it under a context that types its environment: \m{\mathcal{A} \vdash \thunk{\subst}{e} : \tyInst} if there is a \m{\mathrm{\Theta}} with the domain of \m{\subst} such that \m{\mathcal{A}; \mathrm{\Theta} \vdash e : \tyInst} and \m{\mathcal{A} \vdash \subst(x) : \mathrm{\Theta}(x)} for every \m{x}.
In rule \m{[Emb]}, \m{x : \tyInst} says that \m{x} has type \m{\tyInst} in the program.
\end{definition}

\begin{definition}[Data type]
\label{def:datatype}
A type \m{\tyInst} is a \emph{data type} if \m{head(\tyInst)} is \m{(\rightarrow)}, or a type constructor that the program declares as a data type and that is not in \m{emb(\mathcal{A})}.
\end{definition}

\mypara{Proof sketches}
\textsc{Proof sketch of \Cref{thm:embed-typing}.}
First, if \m{\mathrm{\Theta} \vdash e : \tyInst} in the program, then \m{\mathcal{A}; \mathrm{\Theta} \vdash invalidate_{\mathcal{A}}(e) : \tyInst}, by induction on the typing derivation.
A stale constructor, coercion or cast becomes \m{\ub}, which has every type.
Every other coercion is well formed in the embedded signature by \Cref{def:stale}, with the same type, and patterns still type check because the constructors of embedded types stay in the signature.
Second, the theorem follows by induction on the thunk, whose environments are finite because \core has no recursive bindings.
An embedded variable \m{x : \tyInst} is bound to \m{invalidate_{\mathcal{A}}(\rho(x))}, which has type \m{\tyInst} by \m{[Emb]} and the first step; since embedded variables are top level and never rebound, this is the type that the context expects.
\hfill$\square$

\smallskip
\noindent
\textsc{Proof sketch of \Cref{thm:embed-consistent}.}
We follow the consistency proofs for \FC with roles.
Define a rewrite relation on types: an axiom of \m{\mathcal{A}}, or of a \lstinline{newtype} that \m{\mathcal{A}} does not embed, rewrites an instance of its left-hand side to the corresponding right-hand side, at the root or beneath a representational parameter, but never beneath a nominal one.
The relation need not terminate---\m{[\alpha] \mapsto \app{SMT.Seq}{\alpha}} and \m{\app{SMT.Seq}{\alpha} \mapsto [\alpha]} pass \m{[Coh]}---and a left-hand side may repeat a variable, so we argue confluence with parallel reduction rather than through termination.
No two rules overlap: for axioms of \m{\mathcal{A}} by \m{[Coh]}, and an embedded \lstinline{newtype} has no axiom.
No rule rewrites the arguments of another, because the parameters of an embedded type constructor are nominal and \core has no nominal axioms, so nominal equality is syntactic.
Parallel reduction thus has the diamond property, and the relation is confluent.
Next, by induction on \m{\coInst}, two types related by a closed coercion that is not stale rewrite to a common type: \m{sym} and transitivity use confluence, and \m{left} and \m{right} apply only to nominal coercions, which relate syntactically equal types.
Finally, no rule applies at the head of a data type, so rewriting keeps its head, and two data types with a common rewrite share their heads.
\hfill$\square$

\section{The Embeddable Typeclass}
\label{app:embeddable}
This appendix describes how \tool relaxes kind equality for type embeddings (\Cref{sec:impl:embed}).

\mypara{Kind Equality}
Inhabited types in Haskell have kind \lstinline{TYPE rep}.
The representation parameter \lstinline{rep} for almost all types is \lstinline{LiftedRep}: an inhabitant is represented by a pointer at runtime and may return bottom.
For example, \lstinline{SMT.BitVec 64} has kind \lstinline{TYPE LiftedRep}.
Primitive types form an exception.
For example, type \lstinline{Int64#} has kind \lstinline{TYPE Int64Rep}: an
inhabitant is represented as a raw 64-bit value.
Type embeddings require the kinds of their types to be equivalent.
\tool relaxes this constraint to permit embeddings of primitive types: the runtime representation \lstinline{rep} in \lstinline{TYPE rep} may differ.
\tool permits this by instantiating coercion axiom between runtime representations, such as \m{C_{Int64Rep} : Int64Rep \sim LiftedRep}.
%
%
This is coercion evidence is only valid locally for the purpose of symbolic
execution.

\mypara{Coercible}
Haskell does not directly expose syntax to construct a cast.
It does however expose the function \lstinline[columns=fixed]{coerce :: Coercible a b => a -> b} which internally casts its argument into a different type.
The typeclass \lstinline{Coercible} carries a coercion.
Haskell automatically resolves the \lstinline{Coercible} constraint, if possible.
\lstinline{Coercible} fixes the kinds of its type arguments \lstinline{a} and \lstinline{b} to be the same.
For this reason, we cannot use \lstinline{Coercible} for type embeddings such as \lstinline{Int64#} $\mapsto$ \lstinline{SMT.BitVec 64} directly.

\mypara{Embeddable}
We define typeclass \lstinline{Embeddable} that permits a coercion between types with different kinds, up to runtime representation:
\begin{lstlisting}
class (Coercible a (Cast b k1), Cast b k1 ~~ b)
   => Embeddable (a :: k1) (b :: k2)
  type Cast b k1 :: k1
\end{lstlisting}
Haskell rejects definition \lstinline{Coercible a b} due to the kind mismatch between \lstinline{a} and \lstinline{b}.
We create an instance of \lstinline{b} with the correct kind through type family \lstinline{Cast}, which performs cast \m{\cast{b}{\coInst : k2 \sim k1}} internally.
We can require an instance of \lstinline{Coercible} between \lstinline{Cast b k1} and \lstinline{a} as they have the same kind.
The constraint \lstinline{Cast b k1 ~~ b} says that the types are \emph{exactly the same}, but permits their kinds to differ in name.
Otherwise, Haskell has no knowledge \lstinline{Cast} actually performs a cast.
We use the transitive property of coercions: we can coerce from type \lstinline{a} to \lstinline{b} through \lstinline{Cast b k1}.

\section{Case Study}
\subsection{\clash}
\label{app:clash}
\clash is a hardware description library that allows us to write circuits within Haskell.
To support a translation from Haskell to other hardware description languages
such like Verilog, \clash exposes hardware level primitives, like the bit vector
primitive \lstinline{Clash.BitVec n}.
To allow simulating circuits in Haskell, these bit vector types are implemented
using the unbounded integer type \lstinline{Integer}, but to faithfully
represent the generated hardware, we need to model them as \lstinline{SMT.BitVec n}, 
directly.
%
%
In this case study, we show how to write bit vector embeddings for
\lstinline{Clash.BitVec n}.
%

\mypara{Zero Sized Bit Vectors}
The SMT theory of arrays does not support zero sized bit vectors.
This constraint shows up in the type signature of \lstinline{SMT.BitVec}
operations as the typeclass \lstinline{1} \m{\leq} \lstinline{n}.
\begin{lstlisting}
SMT.bvadd :: 1 %$\leq$% n => SMT.BitVec n -> SMT.BitVec n -> SMT.BitVec n
\end{lstlisting}
On the other hand, \lstinline{Clash.BitVec} supports all bit vector sizes.
%
%
\begin{lstlisting}
Clash.bvadd :: Clash.BitVec n -> Clash.BitVec n -> Clash.BitVec n
\end{lstlisting}
If we try to write this embedding as is, Haskell will complain about the missing
instance \lstinline{1} \m{\leq} \lstinline{n}.
\begin{lstlisting}
bvaddE :: Embeddable SMT.BitVec Clash.BitVec
       => Clash.BitVec n -> Clash.BitVec n -> Clash.BitVec n
bvaddE = coerce SMT.bvadd
                ^^^^^^^^^ -- missing constraint 1 %$\leq$% n
\end{lstlisting}
Indeed, we cannot guarantee that we did not receive a zero sized bitvector in this call.
The rejection is proper: \tool would crash upon reaching a zero sized addition,
if this definition were accepted.

\mypara{Wrapper Bit Vector}
To overcome this, we define a wrapper bit vector data type \lstinline{Wrap.BitVec} which is either a zero sized bit vector, or has a positive size and thus carries an \lstinline{SMT.BitVec}.
\begin{lstlisting}
data Wrap.BitVec n where
  BitVecZ :: Wrap.BitVec 0
  BitVecP :: 1 %$\leq$% n => SMT.BitVec n -> Wrap.BitVec n
\end{lstlisting}
The instances of \lstinline{Wrap.BitVec} are mutually exclusive.
The data constructor \lstinline{BitVecZ} carries coercion evidence \lstinline{n} \m{\ncoercion} \lstinline{0}, which says that \lstinline{n} is exactly equivalent to \lstinline{0}.
The data constructor \lstinline{BitVecP} carries the evidence that \lstinline{1} \m{\leq} \lstinline{n}, together with a primitive SMT bit vector.
We can define addition for this type without constraining the bit vector size.
\begin{lstlisting}
Wrap.bvadd :: Wrap.BitVec n -> Wrap.BitVec n -> Wrap.BitVec n
Wrap.bvadd BitVecZ BitVecZ = BitVecZ
Wrap.bvadd (BitVecP lhs) (BitVecP rhs) = BitVecP (SMT.bvadd lhs rhs)
\end{lstlisting}
With this, we instead want to embed our original \clash type \lstinline{Clash.BitVec} as the wrapped instance \lstinline{Wrap.BitVec}.
We can define the term embedding for \lstinline{Clash.bvadd} using \lstinline{Wrap.BitVec}.
\begin{lstlisting}
bvaddE :: Embeddable Wrap.BitVec Clash.BitVec
       => Clash.BitVec n -> Clash.BitVec n -> Clash.BitVec n
bvaddE = coerce Wrap.bvadd
\end{lstlisting}

\mypara{Embedding}
We gather the embeddings in the data type \lstinline{Embedding} which may be passed \tool whenever it is invoked.
\begin{lstlisting}
embeddingBitVec :: Embedding
embeddingBitVec = Embedding
  { typeEmbeddings = [ Clash.BitVec %$\mapsto$% Wrap.BitVec ]
  , termEmbeddings = [ Clash.bvadd %$\mapsto$% bvaddE, .. ] }
\end{lstlisting}
Of course, we need to extend the embedding with more bit vector operations, e.g.
multiplication, concatenation, selects, shifts, etc.
Luckily, these are rather straightforward to provide using the type \lstinline{Wrap.BitVec}.
With all of these operations embedded, we are able to symbolically evaluate \clash circuits.

\subsection{Raw Pointers}
\label{app:ptr}
The \lstinline{IORef} embedding of \cref{sec:case-study} models a single mutable cell: every read or write
addresses the entire referenced value, and the reference itself carries no
arithmetic.
This is enough for \lstinline{IORef}, but not for libraries like \texttt{text} and
\texttt{bytestring} that are implemented directly on top of \lstinline{Ptr} and
\lstinline{ForeignPtr}: raw, byte-addressable pointers that support arithmetic
(e.g. \lstinline{plusPtr}, \lstinline{minusPtr} and \lstinline{castPtr}) and single-byte
access (e.g. \lstinline{peek} and \lstinline{poke}).

\noindent
Instead, we model an array allocation as a tuple of its size and its content, respectively via the word-sized bitvector \lstinline{SMT.BitVec WordSize} and the SMT array primitive \lstinline{SMT.Array (SMT.BitVec WordSize) (SMT.BitVec 8)}.
Array allocations are unique and non-overlapping, just like the references built via our model of \lstinline{IORef}.
In fact, we will reuse the same \lstinline{ModelWorld} for our allocations to build a disjointed map of symbolic arrays, similar to ~\cite{SurveySymExec-CSUR18}.
The \lstinline{Any} contained in the heap of our world is coerced to an instance of the array tuple whenever a deference of \lstinline{Ptr Int8} occurs.
As the array content itself is modeled via the SMT theory of arrays, we are permitted symbolic indexing: the alias analysis is performed by an SMT solver.

\mypara{Pointer Representation}
We embed \lstinline{Ptr a} as a tuple of the allocation identifier into the \lstinline{ModelWorld} that matches a symbolic array and an offset into the symbolic array itself.
\begin{lstlisting}
data ModelPtr a = ModelPtr
  { ptrId  :: Int
  , ptrOff :: SMT.BitVec WordSize
  }
\end{lstlisting}
We embed \lstinline{ForeignPtr a} as just its \lstinline{ptrId}.
The offset is only assigned once \lstinline{withForeignPtr} materializes a
concrete \lstinline{Ptr} at offset 0.

\mypara{Allocation}
Allocation follows exactly the same bump-allocator pattern as
\lstinline{newIORefE}: fetch a fresh id by taking the length of the current
heap, and insert the new entry under that id.
The only difference from \lstinline{newIORefE} is what gets inserted: instead of
the initial value coerced to \lstinline{Any}, we insert a zero-initialized,
constant SMT byte array, itself coerced to \lstinline{Any}:
\begin{lstlisting}
mallocByteStringE :: Embeddable (ForeignPtr a) (ModelForeignPtr a)
                  => Embeddable RealWorld# ModelWorld
                  -> Int -> IO (ForeignPtr a)
mallocByteStringE n = coerce \heap0 ->
  let newId = length heap0
      zeroArray :: SMT.Array (SMT.BitVec WordSize) (SMT.BitVec 8)
      zeroArray = SMT.constArray 0 
      alloc = (n, zeroArray)
      heap1 = insert newId (toAny alloc) heap0
  in (heap1, ModelForeignPtr { ptrId = newId })
\end{lstlisting}
No separate allocator, counter, or heap type is needed to support pointers: an
allocation is an \lstinline{Any} in the same heap \lstinline{IORef} already
uses, and the only new ingredient is the SMT array stored inside.

\mypara{Arithmetic}
\lstinline{plusPtr}, \lstinline{minusPtr}, and \lstinline{castPtr} never touch the
heap at all: they are pure bit-vector operations on \lstinline{ptrOff}, or, for
\lstinline{castPtr}, a pure retype with no runtime effect whatsoever.
\begin{lstlisting}
plusPtrE :: Embeddable (Ptr a) (ModelPtr a)
         => Embeddable Int (SMT.BitVec WordSize)
         => Embeddable (Ptr b) (ModelPtr b)
         => Ptr a -> Int -> Ptr b
plusPtrE = coerce \p n -> p { ptrOff = ptrOff p + n }
\end{lstlisting}
Because \lstinline{n} may be symbolic, so is the resulting \lstinline{ptrOff}: no
concretization ever occurs, which is exactly what lets \tool verify properties over
symbolic offsets, such as that writes at two distinct symbolic offsets do not alias
(\cref{sec:evaluation}).

\mypara{Read and Write}
\lstinline{peek} and \lstinline{poke} first \lstinline{lookup} the allocation's
byte array out of the heap by \lstinline{ptrId}, just as
\lstinline{readIORefE} looks up its value, then \lstinline{unsafeCoerce} the
result back to an SMT array; only then do they \lstinline{SMT.select}/
\lstinline{SMT.store} the byte at \lstinline{ptrOff}, and \lstinline{insert} the
updated array back under the same id:
\begin{lstlisting}
peekByteE :: Embeddable (Ptr Word8) (ModelPtr Word8)
          => Embeddable RealWorld# ModelWorld
          => Ptr Word8 -> IO Word8
peekByteE = coerce \p heap ->
  let (size, alloc) = lookup (ptrId p) heap
      bytes :: SMT.Array (SMT.BitVec WordSize) (SMT.BitVec 8)
      bytes = unsafeCoerce alloc; idx = ptrOff p
      deref = if idx %$\geq$% size then SymFC.UB else SMT.select bytes idx
  in (heap, deref)

pokeByteE :: Embeddable (Ptr Word8) (ModelPtr Word8)
          => Embeddable RealWorld# ModelWorld
          => Ptr Word8 -> Word8 -> IO ()
pokeByteE = coerce \p v heap0 ->
  let (size, alloc) = lookup (ptrId p) heap
      bytes0 :: SMT.Array (SMT.BitVec WordSize) (SMT.BitVec 8)
      bytes0 = unsafeCoerce alloc; idx = ptrOff p
      bytes1 = if idx %$\geq$% size then SymFC.UB else SMT.store bytes0 idx v
      heap1 = insert (ptrId p) (toAny bytes1) heap
  in (heap1, ())
\end{lstlisting}
As in \lstinline{readIORefE}, the \lstinline{unsafeCoerce} back to an SMT array
is safe only because every id ever looked up here was inserted by
\lstinline{mallocByteStringE}, which always stores a value of that same array
type: the same per-id type invariant that already justifies \lstinline{IORef}'s
own usage of \lstinline{unsafeCoerce}.
The \lstinline{ptrOff} may be symbolic here, so aliasing questions \emph{within} a
single allocation, such as whether a write through one pointer affects a read
through another at a possibly-different offset, are answered by the SMT
solver's own array axioms rather than by any logic in the embedding itself.
If the pointer read or write is out of bounds of the allocation, we return \lstinline{SymFC.UB} which models undefined behaviour as described in \cref{sec:theory}.
Aliasing \emph{across} distinct allocations, meanwhile, is decided the same way
it already was for \lstinline{IORef}: two different, concrete ids name two
different heap entries.

\mypara{Embedding}
To summarize, we need the following embeddings for allocations and dereferences.
\begin{lstlisting}
embeddingPtr :: Embedding
embeddingPtr = Embedding
  { typeEmbeddings = [ Ptr a %$\mapsto$% ModelPtr a
                     , ForeignPtr a %$\mapsto$% ModelForeignPtr a ]
  , termEmbeddings = [ plusPtr %$\mapsto$% plusPtrE
                     , mallocForeignPtrBytes %$\mapsto$% mallocByteStringE
                     , withForeignPtr %$\mapsto$% withForeignPtrE
                     , peekByte %$\mapsto$% peekByteE
                     , pokeByte %$\mapsto$% pokeByteE, %$\ldots$% ] }
\end{lstlisting}
Unlike \lstinline{embeddingIORef}, whose heap entries never contain anything the
solver needs to look inside, \lstinline{embeddingPtr} reuses that same heap
outright and pushes only the \emph{contents} of a pointer entry into SMT, not
the heap itself.
The price is a heavier per-allocation representation; the payoff is that
pointer arithmetic and byte-level memory access compose under full symbolic
reasoning exactly where it is needed, which is exactly the bit-level,
pointer-aware modeling that libraries like \texttt{text} and
\texttt{bytestring} require.
This cost is easy to justify in practice, since raw pointer use in Haskell code
is typically confined to a small, contained scope rather than pervasive
throughout a codebase.

\section{Verified Properties and G2Q Benchmarks}
\label{app:properties}
\label{app:g2q}
\Cref{tab:properties-full} lists every property summarised in \Cref{tab:properties}, and \Cref{tab:g2q-full} lists every benchmark summarised in \Cref{tab:g2q}.

\begin{table}[!ht]
\centering\footnotesize
\caption{All G2Q benchmarks. Times are in seconds.}
\label{tab:g2q-full}
\begin{tabular}{lrrr}
\hline
\m{Task}                                               & G2 (s)  & \tool (s) & Grisette (s) \\
\hline
\texttt{badEnvSearch prog}                             & 59.32   & 0.09      & 0.02 \\
\texttt{Search for non-zero Mul x y == Add x y}        & 0.56    & 0.21      & 0.10 \\
\texttt{solveDeBruijn for id}                          & 0.04    & 0.07      & 0.04 \\
\texttt{solveDeBruijn for const}                       & 1.06    & 0.08      & 0.02 \\
\texttt{solveDeBruijn for NOT}                         & Timeout & Timeout   & Timeout \\
\texttt{solveDeBruijn for OR}                          & 86.22   & 0.27      & 0.03 \\
\texttt{solveDeBruijn for AND}                         & Timeout & 20.52     & 12.32 \\
\texttt{solveQueens 4}                                 & 0.36    & 0.08      & 0.03 \\
\texttt{solveQueens 5}                                 & 0.55    & 0.11      & 0.03\\
\texttt{solveQueens 6}                                 & 0.90    & 0.12      & 0.03\\
\texttt{solveQueens 7}                                 & 1.47    & 0.14      & 0.03\\
\texttt{solveQueens 8}                                 & 2.30    & 0.16      & 0.04\\
\texttt{stringSearch for \m{(a+b){*}c(d+(ef)*)}}       & 0.26    & 0.10      & 0.03 \\
\texttt{stringSearch for \m{abcdef}}                   & 4.23    & 0.12      & 0.12 \\
\texttt{stringSearch for \m{a+b+c+d+e+f}}              & 0.05    & 0.09      & 0.04 \\
\texttt{stringSearch for \m{a{*}b{*}c{*}d{*}e{*}f{*}}} & 0.02    & 0.09      & 0.02 \\
\hline
\end{tabular}
\end{table}

\begin{table}[p]
\centering\scriptsize
\caption{%
  All properties verified by \sys with Z3 solve times.
  \cmark\ = valid (no counterexample exists);
  \xmark\ = invalid (counterexample found).
  Times shown are SMT solve times in milliseconds as reported by Z3.
}
\label{tab:properties-full}
\begin{tabular}{@{}p{7.8cm}lcc@{}}
\toprule
Property & Inputs & & ms \\
\midrule
\multicolumn{4}{@{}l}{\textit{Bool}} \\
De Morgan's law & $\m{Bool}^2$ & \cmark & 25 \\
$a \Rightarrow b$ is not a tautology & $\m{Bool}^2$ & \xmark & 8 \\
\addlinespace
\multicolumn{4}{@{}l}{\textit{Integer (unbounded)}} \\
$x + y = y + x$ & $\m{Integer}^2$ & \cmark & 5 \\
$x < x + 1$ & $\m{Integer}$ & \cmark & 6 \\
\addlinespace
\multicolumn{4}{@{}l}{\textit{Int, Int8, Int16, Int32, Int64}} \\
$x + y = y + x$ \textnormal{(one property per type)} & 2 args & \cmark & 5--10 \\
$x + 0 = x$ & $\m{Int}$ & \cmark & 5 \\
$x - x = 0$ & $\m{Int}$ & \cmark & 6 \\
$x \times y = y \times x$ & $\m{Int}^2$ & \cmark & 5 \\
$x < x$ is always false \textnormal{(one property per type)} & 1 arg & \xmark & 5--6 \\
\addlinespace
\multicolumn{4}{@{}l}{\textit{Word, Word8, Word64}} \\
$x + y = y + x$ \textnormal{(one property per type)} & 2 args & \cmark & 6--8 \\
$x + 0 = x$ & $\m{Word}$ & \cmark & 9 \\
$x \mathbin{\&} y = y \mathbin{\&} x$ & $\m{Word}^2$ & \cmark & 5 \\
$x < x$ is always false \textnormal{(one property per type)} & 1 arg & \xmark & 5--6 \\
\addlinespace
\multicolumn{4}{@{}l}{\textit{containers (Data.Map.Strict)}} \\
$\m{member}\ k\ \m{empty} = \m{False}$ & $\m{Int}$ & \cmark & 12 \\
$\m{member}\ k\ (\m{singleton}\ k\ v) = \m{True}$ & $\m{Int}^2$ & \cmark & 5 \\
$\m{lookup}\ k\ (\m{singleton}\ k\ v) = \m{Just}\ v$ & $\m{Int}^2$ & \cmark & 5 \\
$\m{member}\ k\ (\m{insert}\ k\ v\ \m{empty}) = \m{True}$ & $\m{Int}^2$ & \cmark & 9 \\
$\m{lookup}\ k\ (\m{insert}\ k\ v\ \m{empty}) = \m{Just}\ v$ & $\m{Int}^2$ & \cmark & 5 \\
$\m{null}\ (\m{delete}\ k\ (\m{singleton}\ k\ v)) = \m{True}$ & $\m{Int}^2$ & \cmark & 5 \\
$k_1 \neq k_2 \Rightarrow \m{lookup}\ k_1\ (\m{singleton}\ k_2\ v) = \m{Nothing}$ & $\m{Int}^3$ & \cmark & 5 \\
$\m{member}\ k_1\ (\m{singleton}\ k_2\ v)$ without $k_1 = k_2$ precondition & $\m{Int}^3$ & \xmark & 5 \\
$\m{member}\ k\ (\m{delete}\ k\ (\m{singleton}\ k\ v))$ & $\m{Int}^2$ & \xmark & 10 \\
$\m{lookup}\ k\ (\m{insert}\ k\ v_1\ (\m{singleton}\ k\ v_2)) = \m{Nothing}$ & $\m{Int}^3$ & \xmark & 5 \\
\addlinespace
\multicolumn{4}{@{}l}{\textit{containers (Data.Set)}} \\
$\m{member}\ k\ \m{empty} = \m{False}$ & $\m{Int}$ & \cmark & 5 \\
$\m{member}\ k\ (\m{singleton}\ k) = \m{True}$ & $\m{Int}$ & \cmark & 5 \\
$\m{member}\ k\ (\m{insert}\ k\ \m{empty}) = \m{True}$ & $\m{Int}$ & \cmark & 5 \\
$\neg\,\m{member}\ k\ (\m{delete}\ k\ (\m{singleton}\ k))$ & $\m{Int}$ & \cmark & 4 \\
$\m{member}\ k\ (\m{delete}\ k\ (\m{singleton}\ k))$ & $\m{Int}$ & \xmark & 5 \\
\addlinespace
\multicolumn{4}{@{}l}{\textit{iproute (Data.IP)}} \\
$\m{fixByteOrder}\ (\m{fixByteOrder}\ w) = w$ & $\m{Word32}$ & \cmark & 17 \\
IPv4: $a \in \m{makeAddrRange}\ a\ l$ & $\m{Word32},\m{Int}$ & \cmark & 10 \\
IPv4: $r \mathbin{\m{>:>}} r$ & $\m{Word32},\m{Int}$ & \cmark & 5 \\
IPv4: $r_1 \mathbin{\m{>:>}} r_2 \wedge r_2 \mathbin{\m{>:>}} r_3 \Rightarrow r_1 \mathbin{\m{>:>}} r_3$ & $\m{Word32}^3, \m{Int}^3$ & \cmark & 14 \\
IPv6: $a \in \m{makeAddrRange}\ a\ l$ & $\m{Word32}^4, \m{Word8}$ & \cmark & 17 \\
IPv6: $r \mathbin{\m{>:>}} r$ & $\m{Word32}^4, \m{Word8}$ & \cmark & 6 \\
IPv4-mapped address is in lifted IPv6 range & $\m{Word32}, \m{Word8}$ & \cmark & 6 \\
\addlinespace
\multicolumn{4}{@{}l}{\textit{IORef}} \\
Write then read returns the same value & $\m{Int}$ & \cmark & 7 \\
\addlinespace
\multicolumn{4}{@{}l}{\textit{Ptr / ForeignPtr}} \\
For fresh $p$: $\m{peekByte}\ p = 0$ & --- & \cmark & 6 \\
For fresh $p$: $m \neq n \Rightarrow \m{pokeByte}\ (\m{plusPtr}\ p\ m)\ v \mathbin{;} \m{peekByte}\ (\m{plusPtr}\ p\ n) = 0$ & $\m{Int}^2,\m{Word8}$ & \cmark & 7 \\
$\m{pokeByte}\ p\ v_1 \mathbin{;} \m{pokeByte}\ p\ v_2 \mathbin{;} \m{peekByte}\ p = v_2$ & $\m{Word8}^2$ & \cmark & 7 \\
$\m{pokeByte}\ (\m{plusPtr}\ (\m{plusPtr}\ p\ m)\ n)\ v \mathbin{;} \m{peekByte}\ (\m{plusPtr}\ p\ (m+n)) = v$ & $\m{Int}^2,\m{Word8}$ & \cmark & 7 \\
$\m{pokeByte}\ p\ v \mathbin{;} \m{peekByte}\ (\m{castPtr}\ (\m{castPtr}\ p)) = v$ & $\m{Word8}$ & \cmark & 7 \\
$\m{pokeByte}\ (\m{plusPtr}\ p_1\ k)\ v \mathbin{;} \m{peekByte}\ (\m{plusPtr}\ p_2\ k) = 0$ & $\m{Int},\m{Word8}$ & \cmark & 7 \\
\midrule
Total (54 properties) & & $41\cmark,\ 13\xmark$ & 389 \\
\bottomrule
\end{tabular}
\end{table}

\end{document}